\documentclass[runningheads]{llncs}

\usepackage[utf8]{inputenc}
\usepackage{amsmath}
\usepackage{amssymb}
\usepackage{amsfonts}
\usepackage[noabbrev,capitalize]{cleveref}
\usepackage{booktabs}
\usepackage{multirow}
\usepackage{pifont}

\newcommand{\calB}{\ensuremath{\mathcal{B}}}
\newcommand{\calC}{\ensuremath{\mathcal{C}}}
\newcommand{\calF}{\ensuremath{\mathcal{F}}}
\newcommand{\calG}{\ensuremath{\mathcal{G}}}
\newcommand{\calI}{\ensuremath{\mathcal{I}}}
\newcommand{\calJ}{\ensuremath{\mathcal{J}}}
\newcommand{\calK}{\ensuremath{\mathcal{K}}}
\newcommand{\calO}{\ensuremath{\mathcal{O}}}
\newcommand{\calP}{\ensuremath{\mathcal{P}}}
\newcommand{\calT}{\ensuremath{\mathcal{T}}}
\newcommand{\N}{\ensuremath{\mathbb{N}}}

\newcommand\qedblob{\ding{113}}
\def\literalqed{{\ \nolinebreak\hfill\mbox{\qedblob\quad}}}

\def\qed{\literalqed}

\newenvironment{proofs}{\noindent{\bf Proof.}\hspace*{1em}}{\literalqed\bigskip}
\newcommand{\sproofof}[1]{\noindent{\bf Proof of {#1}.}\hspace*{1em}}

\newcommand{\eproofof}[1]{\hfill \mbox{\qed\quad$_{\mbox{\small {#1}}}$}\bigskip}

\newcommand{\Pol}{\ensuremath{\mathrm{P}}}
\newcommand{\NP}{\ensuremath{\mathrm{NP}}}
\newcommand{\DP}{\ensuremath{\mathrm{DP}}}
\newcommand{\ThetaZP}{\ensuremath{\Theta_{2}^{\Pol}}}

\newcommand{\stability}{\ensuremath{\textsc{Stability}}}

\newcommand{\vstability}{\ensuremath{\textsc{VertexStability}}}
\newcommand{\vcriticality}{\ensuremath{\textsc{VertexCriticality}}}
\newcommand{\unfrozenness}{\ensuremath{\textsc{Unfrozenness}}}
\newcommand{\frozenness}{\ensuremath{\textsc{Frozenness}}}
\newcommand{\vunfrozenness}{\ensuremath{\textsc{VertexUnfrozenness}}}

\newtheorem{observation}{Observation}

\usepackage[
  colorinlistoftodos,
  prependcaption,
  textsize = small,
]{todonotes}
\definecolor{asparagus}{rgb}{0.53, 0.66, 0.42}

\begin{document}
\title{Stability of Special Graph Classes
}
\author{Robin Weishaupt
  \and
  J\"org Rothe}
\authorrunning{R. Weishaupt and J. Rothe}
\institute{Institut f\"ur Informatik,
  Heinrich-Heine-Universit\"at D\"usseldorf,
  D\"usseldorf, Germany\\
  \email{robin.weishaupt@hhu.de, rothe@hhu.de}\\
}
\maketitle
\begin{abstract}
Frei et al.~\cite{fre-hem-rot:c:complexity-of-stability} showed that
the problem to decide whether a graph is stable with respect to some
graph parameter under adding or removing either edges or vertices is
$\ThetaZP$-complete.  They studied the common graph parameters
$\alpha$ (independence number), $\beta$ (vertex cover number),
$\omega$ (clique number), and $\chi$ (chromatic number) for certain
variants of the stability problem.  We follow their approach and
provide a large number of
polynomial-time algorithms solving these problems for special graph
classes, namely for graphs without edges, complete graphs, paths,
trees, forests, bipartite graphs, and co-graphs.
\end{abstract}

\section{Introduction}
\label{sec:introduction}

Frei et al.~\cite{fre-hem-rot:c:complexity-of-stability}
comprehensively studied the problem of how stable certain central
graph parameters are when a given graph is slightly modified, i.e.,
under operations such as adding or deleting either edges or vertices.
Given a graph parameter $\xi$ (like, e.g., the independence number or
the chromatic number), they formally introduced the problems
$\xi$-$\stability$, $\xi$-$\vstability$, $\xi$-$\unfrozenness$, and
$\xi$-$\vunfrozenness$ and showed that they are, typically,
$\ThetaZP$-complete, that is, they are complete for the complexity
class known as ``parallel access to $\NP$,'' which was introduced by
Papadimitriou and Zachos~\cite{pap-zac:c:two-remarks} and intensely
studied by, e.g., Wagner~\cite{wag:j:min-max,wag:j:bounded},
Hemaspaandra et al.~\cite{hem-hem-rot:j:dodgson,hem-spa-vog:j:kemeny},
and Rothe et al.~\cite{rot-spa-vog:j:young}; see the survey by
Hemaspaandra et
al.~\cite{hem-hem-rot:j:raising-lower-bounds-survey}.\footnote{$\ThetaZP$
  is contained in the second level of the polynomial hierarchy and
  contains the problems that can be solved in polynomial time by an
  algorithm that accesses its $\NP$ oracle in parallel (i.e., it first
  computes all its queries and then asks them all at once and accepts
  its input depending on the answer vector).  Alternatively,
  $\ThetaZP=\Pol^{\NP[\mathcal{O}(\log n)]}$ can be viewed as the
  class of problems solvable in polynomial time via \emph{adaptively}
  accessing its $\NP$ oracle (i.e., computing the next query depending
  on the answer to the previous query) logarithmically often (in the
  input size).}

Furthermore, Frei et al.~\cite{fre-hem-rot:c:complexity-of-stability}
proved that some more specific versions of these problems, namely
$k$-$\chi$-$\stability$ and $k$-$\chi$-$\vstability$, are
$\NP$-complete for $k = 3$ and $\DP$-complete for $k \geq 4$,
respectively, where $\DP$ was introduced by Papadimitriou and
Yannakakis~\cite{pap-yan:j:dp} as the class of problems that can
be written as the difference of $\NP$ problems.

Overall, the results of Frei et
al.~\cite{fre-hem-rot:c:complexity-of-stability} indicate that these
problems are rather intractable and there exist no efficient
algorithms
solving them exactly.
Considering the vast number of real-world applications for these
problems mentioned by
Frei et al.~\cite{fre-hem-rot:c:complexity-of-stability}---e.g., the
design of infrastructure, coloring algorithms for biological
networks~\cite{min-urb:j:graph-theory-landscape-connectivity-conservation,kho:j:application-of-graph-coloring-to-biological-networks}
or for complex information, social, and economic
networks~\cite{jac:b:social-and-economic-networks}, etc.---these
results are rather disappointing and unsatisfying.

This obstacle motivates us to study whether there are scenarios that
allow for efficient solutions to these problems which in general are
intractable.  Our work is based on the assumption that most of the
real-world applications of stability of graph parameters do not use
arbitrarily complex graphs but
may often be restricted to certain special graph classes.
Consequently, our studies show that---despite the completeness results
by Frei et al.~\cite{fre-hem-rot:c:complexity-of-stability}---there
are tractable solutions to these problems when one limits the scope of
the problem to a special graph class.
We study seven different classes of special graphs: empty graphs
consisting of only isolated vertices and no edges~($\calI$), complete
graphs that have all possible edges~($\calK$), paths ($\calP$), trees
($\calT$), forests ($\calF$), bipartite graphs ($\calB$), and
co-graphs~($\calC$).  For each such class, we study twelve different
problems:
\begin{itemize}
  \item stability, vertex-stability, and unfrozenness
  \item for the four graph parameters~$\alpha$, $\beta$, $\omega$,
    and~$\chi$.
\end{itemize}

In total, we thus obtain the 84 $\Pol$ membership results shown in
Table~\ref{tab:results}, which gives the theorem, proposition, or
corollary corresponding to each such $\Pol$ result.
This can be useful for real-world applications when knowledge about
the stability, vertex-stability, or unfrozenness of a graph with
respect to a certain graph parameter is required and graphs with such
a special structure may typically occur in this application.

\begin{table}[t!]
  \centering
  \caption{Overview of $\Pol$ results established for seven special
    graph classes in this paper. E stands for the edge-related problem
    and V for the vertex-related problem.}
  \begin{tabular}{l r r @{\quad} c c c c c c c}
    \toprule
    &&& $\calI$ & $\calK$ & $\calP$ & $\calT$ & $\calF$ & $\calB$ & $\calC$ \\
    \midrule
    \multirow{3}{*}{$\alpha$} & \multirow{2}{*}{$\stability$} & E
                & Thm.~\ref{thm:empty-graphs-stability-vertexstability}
                & Prop.~\ref{prop:complete-graphs-xi-critical}
                & Thm.~\ref{thm:path-stability-vertexstability}
                & Cor.~\ref{cor:tree-stability-vertexstability}
                & Thm.~\ref{thm:forests-stability-vertexstability}
                & Cor.~\ref{cor:bipartite-stability-vertexstability}
                & Cor.~\ref{cor:co-graph-alpha-stability} \\
           && V & Thm.~\ref{thm:empty-graphs-stability-vertexstability}
                & Cor.~\ref{cor:complete-graphs-vertexstability}
                & Thm.~\ref{thm:path-stability-vertexstability}
                & Cor.~\ref{cor:tree-stability-vertexstability}
                & Thm.~\ref{thm:forests-stability-vertexstability}
                & Cor.~\ref{cor:bipartite-stability-vertexstability}
                & Thm.~\ref{thm:cograph-alpha-vertexstability} \\
           & \multicolumn{2}{r@{\quad}}{$\unfrozenness$}
                & Cor.~\ref{cor:empty-graphs-unfrozenness}
                & Cor.~\ref{cor:complete-graph-unfrozenness}
                & Prop.~\ref{prop:path-alpha-beta-unfrozenness}
                & Cor.~\ref{cor:tree-forest-alpha-beta-unfrozenness}
                & Cor.~\ref{cor:tree-forest-alpha-beta-unfrozenness}
                & Cor.~\ref{cor:bipartite-alpha-unfrozenness}
                & Cor.~\ref{cor:co-graph-alpha-beta-unfrozenness} \\
    \midrule
    \multirow{3}{*}{$\beta$} & \multirow{2}{*}{$\stability$} & E
                & Thm.~\ref{thm:empty-graphs-stability-vertexstability}
                & Prop.~\ref{prop:complete-graphs-xi-critical}
                & Thm.~\ref{thm:path-stability-vertexstability}
                & Cor.~\ref{cor:tree-stability-vertexstability}
                & Thm.~\ref{thm:forests-stability-vertexstability}
                & Cor.~\ref{cor:bipartite-stability-vertexstability}
                & Cor.~\ref{cor:co-graph-beta-stability} \\
           && V & Thm.~\ref{thm:empty-graphs-stability-vertexstability}
                & Cor.~\ref{cor:complete-graphs-vertexstability}
                & Thm.~\ref{thm:path-stability-vertexstability}
                & Cor.~\ref{cor:tree-stability-vertexstability}
                & Thm.~\ref{thm:forests-stability-vertexstability}
                & Cor.~\ref{cor:bipartite-stability-vertexstability}
                & Cor.~\ref{cor:co-graph-beta-vertex-stability} \\
           & \multicolumn{2}{r@{\quad}}{$\unfrozenness$}
                & Cor.~\ref{cor:empty-graphs-unfrozenness}
                & Cor.~\ref{cor:complete-graph-unfrozenness}
                & Prop.~\ref{prop:path-alpha-beta-unfrozenness}
                & Cor.~\ref{cor:tree-forest-alpha-beta-unfrozenness}
                & Cor.~\ref{cor:tree-forest-alpha-beta-unfrozenness}
                & Thm.~\ref{thm:bipartite-beta-unfrozenness}
                & Cor.~\ref{cor:co-graph-alpha-beta-unfrozenness} \\
    \midrule
    \multirow{3}{*}{$\omega$} & \multirow{2}{*}{$\stability$} & E
                & Thm.~\ref{thm:empty-graphs-stability-vertexstability}
                & Prop.~\ref{prop:complete-graphs-xi-critical}
                & Thm.~\ref{thm:path-stability-vertexstability}
                & Cor.~\ref{cor:tree-stability-vertexstability}
                & Thm.~\ref{thm:forests-stability-vertexstability}
                & Cor.~\ref{cor:bipartite-stability-vertexstability}
                & Thm.~\ref{thm:co-graph-omega-stability} \\
           && V & Thm.~\ref{thm:empty-graphs-stability-vertexstability}
                & Cor.~\ref{cor:complete-graphs-vertexstability}
                & Thm.~\ref{thm:path-stability-vertexstability}
                & Cor.~\ref{cor:tree-stability-vertexstability}
                & Thm.~\ref{thm:forests-stability-vertexstability}
                & Cor.~\ref{cor:bipartite-stability-vertexstability}
                & Cor.~\ref{cor:co-graph-omega-vertex-stability} \\
           & \multicolumn{2}{r@{\quad}}{$\unfrozenness$}
                & Cor.~\ref{cor:empty-graphs-unfrozenness}
                & Cor.~\ref{cor:complete-graph-unfrozenness}
                & Prop.~\ref{prop:path-omega-chi-unfrozen}
                & Prop.~\ref{prop:tree-omega-chi-unfrozenness}
                & Thm.~\ref{thm:forest-omega-chi-unfrozenness}
                & Cor.~\ref{cor:bipartite-omega-unfrozenness}
                & Cor.~\ref{cor:co-graph-omega-unfrozenness} \\
    \midrule
    \multirow{3}{*}{$\chi$} & \multirow{2}{*}{$\stability$} & E
                & Thm.~\ref{thm:empty-graphs-stability-vertexstability}
                & Prop.~\ref{prop:complete-graphs-xi-critical}
                & Thm.~\ref{thm:path-stability-vertexstability}
                & Cor.~\ref{cor:tree-stability-vertexstability}
                & Thm.~\ref{thm:forests-stability-vertexstability}
                & Cor.~\ref{cor:bipartite-stability-vertexstability}
                & Thm.~\ref{thm:co-graph-chi-stability} \\
           && V & Thm.~\ref{thm:empty-graphs-stability-vertexstability}
                & Cor.~\ref{cor:complete-graphs-vertexstability}
                & Thm.~\ref{thm:path-stability-vertexstability}
                & Cor.~\ref{cor:tree-stability-vertexstability}
                & Thm.~\ref{thm:forests-stability-vertexstability}
                & Cor.~\ref{cor:bipartite-stability-vertexstability}
                & Thm.~\ref{thm:co-graph-chi-vertex-stability} \\
           & \multicolumn{2}{r@{\quad}}{$\unfrozenness$}
                & Prop.~\ref{prop:empty-graphs-chi-unfrozenness}
                & Cor.~\ref{cor:complete-graph-unfrozenness}
                & Prop.~\ref{prop:path-omega-chi-unfrozen}
                & Prop.~\ref{prop:tree-omega-chi-unfrozenness}
                & Thm.~\ref{thm:forest-omega-chi-unfrozenness}
                & Thm.~\ref{thm:bipartite-chi-unfrozenness}
                & Thm.~\ref{thm:co-graph-chi-unfrozenness} \\
    \bottomrule
  \end{tabular}
  \label{tab:results}
\end{table}

\section{Preliminaries}
\label{sec:preliminaries}

We follow the notation of
Frei et al.~\cite{fre-hem-rot:c:complexity-of-stability} and briefly
collect the relevant notions here (referring to their
paper~\cite{fre-hem-rot:c:complexity-of-stability} for further discussion).
Let $\calG$ be the set of all undirected, simple graphs without loops.
For $G \in \calG$, we denote by $V(G)$ its vertex set and by
$E(G)$ its edge set;
by $\overline{G}$ its complementary graph with $V(\overline{G}) =
V(G)$ and $E(\overline{G}) = \{ \{ v, w \} \in V(G) \times V(G) \mid v
\neq w \wedge \{ v, w \} \notin E(G) \}$.  For $v\in V(G)$, $e\in
E(G)$, and $e'\in E(\overline{G})$, let $G-v$, $G-e$, and $G+e'$,
respectively, denote the graphs that result from $G$ by deleting~$v$,
deleting~$e$, and adding~$e'$.

A \emph{graph parameter} is a map $\xi: \mathcal{G}\to\N$.  We focus on
the prominent graph parameters $\alpha$ (the size of a maximum
independent set), $\beta$ (the size of a minimum vertex cover), $\chi$
(the chromatic number, i.e., the minimum number of colors needed to
color the vertices of a graph so that no two adjacent vertices have
the same color), and $\omega$ (the size of a maximum clique).

For a graph parameter~$\xi$, an edge $e\in E(G)$ is said to be
$\xi$-\emph{stable} if $\xi(G)=\xi(G-e)$, i.e., $\xi(G)$ remains
unchanged after $e$ is deleted from~$G$.  Otherwise (i.e., if $\xi(G)$
is changed by deleting~$e$), $e$ is said to be $\xi$-\emph{critical}.
Stability and criticality are defined analogously for a vertex $v\in
V(G)$ instead of an edge $e\in E(G)$.

A graph is said to be \emph{$\xi$-stable} if all its edges are
$\xi$-stable.  A graph whose vertices (instead of edges) are all
$\xi$-stable is said to be \emph{$\xi$-vertex-stable}, and
\emph{$\xi$-criticality} and \emph{$\xi$-vertex-criticality} are
defined analogously.  Obviously, each edge and each vertex is either
stable or critical, yet a graph might be neither.

Traditionally, the analogous terms for stability or vertex-stability
when an edge or a vertex is \emph{added} rather than deleted are
\emph{unfrozenness} and \emph{vertex-unfrozenness}: They too indicate
that a graph parameter does not change by this operation.  And if,
however, a graph parameter \emph{changes} when an edge or vertex is
\emph{added} (not deleted), the notions analogous to criticality and
vertex-criticality are simply termed \emph{frozenness} and
\emph{vertex-frozenness}.  Again, each edge and each vertex is either
unfrozen or frozen, but a graph might be neither.

For a graph parameter~$\xi$, define $\xi$-$\stability$ to be the set
of $\xi$-stable graphs; and analogously so for the sets
$\xi$-$\vstability$, $\xi$-$\vcriticality$, $\xi$-$\unfrozenness$,
$\xi$-$\frozenness$, and $\xi$-$\vunfrozenness$.  These are the
decision problems studied by Frei et
al.~\cite{fre-hem-rot:c:complexity-of-stability} for general graphs in
terms of their computational complexity.  We will study them
restricted to the
graph classes mentioned in the introduction,
formally defined in the
subsections of
Section~\ref{sec:special-graph-classes}.

A graph class $\calJ \subseteq \calG$ is closed for (induced) subgraphs
if for every $G \in \calJ$ it holds that all (induced) subgraphs $H$
of $G$ satisfy $H \in \calJ$.

The notation of perfect graphs was originally introduced by
Berge~\cite{ber:j:perfect-graphs} in 1963: A graph $G \in \calG$ is
called \emph{perfect} if for all induced subgraphs
$H$ of~$G$, we have $\chi(H) = \omega(H)$.  Note that $G$ is also an
induced subgraph of itself.

Let $\Pol$ be the class of problems solvable in (deterministic)
polynomial time.  For more background on computational complexity
(e.g., regarding the complexity classes $\NP$, $\DP$, and $\ThetaZP$
mentioned in the introduction; note that $\Pol \subseteq \NP \subseteq
\DP \subseteq \ThetaZP$ by definition), we refer to the textbooks by
Papadimitriou~\cite{pap:b:complexity} and
Rothe~\cite{rot:b:cryptocomplexity}.

\section{General Stability and Unfrozenness Results}
\label{sec:general-stability}

In this section, we provide general results that hold for specific
graph classes satisfying special requirements.  These results can be
used to easily determine for a given graph class whether some
stability or unfrozenness results are tractable.

\begin{theorem}
  \label{thm:general-efficient-problems}
  Let $\calJ \subseteq \calG$ be a graph class closed for
  induced subgraphs, and $\xi$ a tractable graph parameter for $\calJ$.
  Then $\xi$-$\textsc{VertexStability} \in \Pol$ for all $G \in \calJ$.
\end{theorem}

\begin{proofs}
  Let $G \in \calJ$ and compute $\xi(G)$. For all $v \in V(G)$, we have
  $G - v \in \calJ$, since $\calJ$ is closed for induced
  subgraphs. Hence, for all $v \in V(G)$, we can compute $\xi(G - v)$
  efficiently and compare it to $\xi(G)$. If there is no vertex such
  that the values differ, $G$ is $\xi$-vertex-stable. This approach is
  computable in time polynomial in $|G|$, so that
  $\xi$-$\textsc{VertexStability} \in \Pol$ for all $G \in \calJ$.
\end{proofs}

Since every graph class that is closed for subgraphs is also closed
for induced subgraphs, Corollary~\ref{cor:induced} is a simple
consequence of the previous theorem.

\begin{corollary}
  \label{cor:induced}
  Let $\calJ \subseteq \calG$ be a graph class closed under
  subgraphs and $\xi$ a tractable graph parameter for $\calJ$. Then
  $\xi$-$\vstability \in \Pol$ for all $G \in \calJ$.
\end{corollary}

The first theorem made a statement related to vertex-stability about
graph classes closed for induced
subgraphs. Theorem~\ref{thm:general-closed-subgraph-stability-in-p} is
related to edge-stability; its proof is deferred to the appendix, as
is the case for most upcoming results.

\begin{theorem}
  \label{thm:general-closed-subgraph-stability-in-p}
  Let $\calJ \subseteq \calG$ be a graph class closed under
  subgraphs and $\xi$ a tractable graph parameter for $\calJ$. Then
  $\xi$-$\textsc{Stability} \in \Pol$ for all $G \in \calJ$.
\end{theorem}

Some of the special graph classes we study in the next section are
perfect,
which is why we now provide some results for perfect graph classes.

\begin{theorem}
  \label{thm:perfect}
  Let $G \in \calG$ be a perfect graph. Then it holds that $G$ is
  $\omega$-vertex-stable if and only if $G$ is $\chi$-vertex-stable.
\end{theorem}

Based on this result, the next corollary follows immediately.

\begin{corollary}
  \label{cor:perfect-graphs-omega-chi}
  Let $\calJ \subseteq \calG$ be a class of perfect graphs. Then, for
  all graphs in $\calJ$, we have $\chi$-$\vstability = \omega$-$\vstability$.
\end{corollary}

While the above results are related to the concepts of stability and
vertex-stability, the subsequent two results address the topic of
unfrozenness.

\begin{theorem}
  \label{thm:general-closed-complement-subgraph-omega-unfrozenness-in-p}
  Let $\calJ \subseteq \calG$ be a graph class closed under
  complements and subgraphs. If $\alpha$ or $\beta$ is tractable for
  $\calJ$, then $\omega$-$\unfrozenness \in \Pol$ for all $G \in \calJ$.
\end{theorem}

Note that this theorem exploits a relation between $\alpha$- and
$\beta$-$\stability$ and $\omega$-$\unfrozenness$.  The next theorem
follows by a similar approach, but exploits a relation between
$\omega$-$\stability$ and $\alpha$- and $\beta$-$\unfrozenness$.

\begin{theorem}
  \label{thm:general-other}
  Let $\calJ \subseteq \calG$ be a graph class closed under
  complements and subgraphs. If $\omega$ is tractable for $\calJ$, then
  $\alpha$- and $\beta$-$\unfrozenness \in \Pol$ for all $G \in \calJ$.
\end{theorem}

\section{Tractability Results for Special Graph Classes}
\label{sec:special-graph-classes}

Ahead of our results for the individual graph classes, we provide two
observations (proven in the appendix) which we will use multiple times in
upcoming proofs.\footnote{Note that the second observation is in line
with Observation~2 of Frei et
al.~\cite{fre-hem-rot:c:complexity-of-stability}.}

\begin{observation}
  \label{obs:vertex-stability-subset-stability}
  $\chi$-$\textsc{VertexStability} \subseteq \chi$-$\textsc{Stability}.$
\end{observation}

\begin{observation}
  \label{beta-critical-edge-critical-vertices}
  Let $G \in \calG$ be a graph. If an edge $\{ u, v \} \in E(G)$ is
  $\beta$-critical, then $u$ and $v$ are $\beta$-critical, too.
\end{observation}

With these two observations we can now inspect several
graph classes. In the following subsections we
study the problems $\xi$-$\stability$, $\xi$-$\vstability$, and
$\xi$-$\unfrozenness$ with $\xi \in \{ \alpha, \beta, \omega, \chi \}$,
restricted to special graph classes.  Frei et
al.~\cite{fre-hem-rot:c:complexity-of-stability} showed that for $\xi
\in \{ \alpha, \omega, \chi \}$ we have $\xi$-$\vunfrozenness =
\emptyset$ as well as $\beta$-$\vunfrozenness = \{ (\emptyset,
\emptyset) \}$, where $(\emptyset, \emptyset)$ is the \emph{null
  graph}, i.e.,  the graph without vertices or edges.

This is why we do not study problems related to vertex-unfrozenness,
as all related questions are already answered.

\subsection{Empty Graphs}
\label{subsec:empty-graphs}

Let $I_n = (\{v_1, \ldots, v_n\}, \emptyset)$ denote the \emph{empty graph}
with $n \in \N$ isolated vertices and $\calI = \{ I_n \mid n \in \N
\}$ the set of all empty graphs.

\begin{theorem}
  \label{thm:empty-graphs-stability-vertexstability}
  It holds that $I_n \in \calI$ is
  (1) $\chi$-vertex-stable for $n = 0$ and $n \geq 2$,
  (2) $\omega$-vertex-stable for $n = 0$ and $n \geq 2$,
  (3) $\beta$-vertex-stable,
  (4) $\alpha$-vertex-stable only for $n = 0$, and
  (5) $\xi$-stable for $\xi \in \{ \alpha, \beta, \chi, \omega \}$.
\end{theorem}

With the previous theorem we can efficiently decide for every empty
graph whether it is $\xi$-stable or $\xi$-vertex-stable for $\xi
\in \{\alpha, \beta, \chi, \omega \}$.

Furthermore, \cref{thm:empty-graphs-stability-vertexstability} also
provides results for the null graph $(\emptyset, \emptyset)$.
Therefore, we
exclude the null graph from all subsections
hereafter in order to decrease redundancy.

\begin{proposition}\label{prop:empty-graphs-chi-unfrozenness}
  The only $\chi$-unfrozen empty graphs are $I_0$ and $I_1$.
\end{proposition}

The next corollary, which follows from the
results in the next subsection, shows that all remaining problems related to
unfrozenness are in $\Pol$, too.

\begin{corollary}\label{cor:empty-graphs-unfrozenness}
  $\alpha$-, $\beta$-, and $\omega$-$\unfrozenness$ belong
  to $\Pol$ for empty graphs.
\end{corollary}

\subsection{Complete Graphs}\label{subsec:complete-graphs}

Since we studied empty graphs, it was an immediate consequence that we
also study complete graphs. A \emph{complete graph} with $n \in \N$ vertices
is defined as $K_n = (\{ v_1, \ldots, v_n \}, \{ \{ v_i, v_j \} \mid 1
\leq i < j \leq n \})$ and we denote the set of all complete graphs by
$\calK$. We know that $\calI \subseteq \beta$-$\vstability$
holds. Together with the first statement of Proposition~6 from
\cite{fre-hem-rot:c:complexity-of-stability} we obtain that $\calK
\subseteq \omega$-$\vcriticality$, so every complete graph is
$\omega$-vertex-critical. Furthermore, we know that $\calI \setminus
\{ I_1 \} \subseteq \omega$-$\vstability$. Applying the second statement of
Proposition~6 from \cite{fre-hem-rot:c:complexity-of-stability},
we obtain that $\calK \setminus \{ K_1 \} \subseteq
\alpha$-$\vstability$ as well as $\calK \setminus \{ K_1 \} \subseteq
\beta$-$\vcriticality$. Consequently, every complete graph which
does not possess exactly one vertex is $\beta$-vertex-critical and
$\alpha$-vertex-stable.

\begin{observation}\label{obs:complete-graphs-chi-vertex-criticality}
  It holds that $\calK \subseteq \chi$-$\vcriticality$.
\end{observation}

So far we know for all parameters $\xi \in \{ \alpha, \beta, \omega,
\chi \}$ whether complete graphs are vertex-stable or not, summarized
in the subsequent corollary.

\begin{corollary}\label{cor:complete-graphs-vertexstability}
  For every $\xi \in \{\alpha, \beta, \omega, \chi\}$, the problem
  $\xi$-$\vstability$ belongs to $\Pol$ for complete graphs.
\end{corollary}

Next, let us take a look at the edge-related stability problems.

\begin{observation}
  For $K_n \in \calK$, we have $\chi(K_n) = n$, $\alpha(K_n) = 1$,
  $\beta(K_n) = n - 1$, and $\omega(K_n) = n$.
\end{observation}

Since $K_0 = I_0$ and $K_1 = I_1$, these cases were already
covered in the previous subsection, so the next result
is stated only for $n \geq 2$.

\begin{proposition}\label{prop:complete-graphs-xi-critical}
  Let $n \in \N$ with $n \geq 2$. Then $K_n$ is $\xi$-critical for
  $\xi \in \{ \alpha, \beta, \omega, \chi \}$.
\end{proposition}

We call a complete graph complete because all possible edges (ignoring
loops) are present in the graph.  Hence, for every $K_n \in \calK$, we
have $\overline{E}(K_n) = \emptyset$, so the next corollary is
immediately clear.\footnote{This corollary is in line with
  \cite[Proposition~5(2)]{fre-hem-rot:c:complexity-of-stability},
  as $\calI = \{ \overline{K_n} \mid n \in \N \}$.}

\begin{corollary}\label{cor:complete-graph-unfrozenness}
  For all $\xi \in \{ \alpha, \beta, \omega, \chi \}$,
  every $K_n \in \calK$ is $\xi$-unfrozen.
\end{corollary}

\subsection{Paths}\label{subsec:paths}

Denote by $P_n = (\{ v_1, \ldots, v_n \}, \{ \{v_i,v_{i+1}\} \mid 1
\leq i < n \})$ the \emph{path} with $n$ vertices and by $\calP$ the set of
all paths. Again, all proofs are deferred to the appendix.

\begin{observation}\label{obs:path-parameters}
  For $n \geq 2$, we have $\chi(P_0) = 0$, $\chi(P_1) = 1$, and
  $\chi(P_n) = 2$ and $\omega(P_0) = 0$, $\omega(P_1) = 1$, and
  $\omega(P_n) = 2$. Additionally, for $n \geq 0$, we have
  $\beta(P_n) = \lfloor \frac{n}{2} \rfloor$ and
  $\alpha(P_n) = \lceil \frac{n}{2} \rceil$.
\end{observation}

Having made this straightforward observation, we can formulate the
following stability results for paths.
Thereby, $P_0 = (\emptyset, \emptyset)$ is ignored,
as argued earlier.

\begin{theorem}\label{thm:path-stability-vertexstability}
   Let $\xi \in \{ \alpha, \beta, \chi, \omega \}$.
  $P_1$ is $\xi$-stable and $\beta$-vertex-stable but not $\rho$-vertex-stable
  for $\rho \in \{ \chi, \alpha, \omega \}$.
  $P_2$ is neither $\xi$-stable nor $\rho$-vertex-stable
  for $\rho \in \{ \beta, \chi, \omega \}$,
  but it is $\alpha$-vertex-stable.
  $P_3$ is $\xi$-stable but not $\xi$-vertex-stable.
  For $n \geq 4$, $P_n$ is $\chi$- and $\omega$-stable
  as well as
  $\chi$- and $\omega$-vertex-stable; it is not $\beta$-vertex-stable;
  and it is neither $\alpha$-stable nor $\beta$-stable but
  $\alpha$-vertex-stable if $n$ is even, and it is $\alpha$-stable and
  $\beta$-stable but not $\alpha$-vertex-stable if $n$ is odd.
\end{theorem}

This theorem yields for all paths $\calP \subseteq \calG$ and $\xi \in
\{ \alpha, \beta, \chi, \omega \}$ that $\xi$-$\textsc{Stability}$ and
$\xi$-$\textsc{VertexStability}$ are in~$\Pol$. We exclude $P_1 = I_1$
onwards.

\begin{observation}\label{obs:path-2-3-unfrozen-frozen}
  For $\xi \in \{ \alpha, \beta, \omega, \chi \}$, $P_2$ is
  $\xi$-unfrozen and $P_3$ is $\xi$-frozen.
\end{observation}

It remains to study the unfrozenness of paths with $n \geq 4$
vertices.

\begin{proposition}\label{prop:path-omega-chi-unfrozen}
  For $n \geq 4$, $P_n$
  is neither $\chi$-unfrozen nor $\omega$-unfrozen.
\end{proposition}

\begin{proposition}\label{prop:path-alpha-beta-unfrozenness}
  For $n \geq 4$, $P_n$ is neither $\alpha$- nor $\beta$-unfrozen if
  $n$ is odd, and $P_n$ is $\alpha$- and $\beta$-unfrozen if $n$ is
  even.
\end{proposition}

Again, $\xi$-$\unfrozenness$ is in~$\Pol$ for all paths $\calP
\subseteq \calG$ and $\xi \in \{ \alpha, \beta, \chi, \omega \}$.

\subsection{Trees and Forests}\label{subsec:trees-forests}

We say $G \in \calG$ is a \emph{tree} (i.e.,
$G \in \calT$) if $G$ has no isolated vertices and no cycles of length
greater than or equal to~$3$. Furthermore, $G$ is a \emph{forest} (i.e.,
$G \in \calF$) if
there exist trees $G_1, \ldots, G_n \in \calT$ such that
$G = G_1 \cup \cdots \cup G_n$.
For every tree $G \in \calT$, it holds that
$|E(G)| = |V(G)| - 1$ (see,
e.g., Bollobás~\cite{bol:b:modern-graph-theory}). So,
we have
$\omega(G) = \chi(G) = 2$ if $|V(G)| > 1$,
and $\omega(G) = \chi(G) = 1$ if $|V(G)| = 1$.

Also, there exists a tractable algorithm to determine
$\alpha(G)$ for trees (for example, as $\calT \subseteq
  \calB$, we can simply use the algorithm for bipartite graphs
  from \cref{obs:bipartite-alpha-beta-efficiently}).
Thus we can compute $\beta$ for trees using Gallai's
theorem~\cite{gal:j:extreme-punkt-kantenmengen} (stated as
Theorem~\ref{thm:gallai} in the appendix), and all four graph
parameters $\alpha$, $\beta$, $\omega$, and $\chi$ are tractable for
trees.

Now, let $G \in \calF$ with $G = G_1 \cup \cdots \cup G_n$
and $G_i \in \calT$, $1 \leq i \leq n$, be a forest.
It is easy to check that
$\alpha(G) = \sum_{i=1}^{n} \alpha(G_i)$,
$\beta(G) = \sum_{i=1}^{n} \beta(G_i)$,
$\omega(G) = \max_{1 \leq i \leq n} \omega(G_i)$, and
$\chi(G) = \max_{1 \leq i \leq n} \chi(G_i)$.
Furthermore, it is known that the class of forests $\calF$ is closed
under subgraphs and induced subgraphs.  From these observations we
have the following results (with proofs in the appendix).

\begin{theorem}
  \label{thm:forests-stability-vertexstability}
  Let $\xi \in \{ \alpha, \beta, \omega, \chi \}$ be a graph
  parameter. Then the problems $\xi$-$\textsc{Stability}$ and
  $\xi$-$\textsc{VertexStability}$ are in $\Pol$ for all forests.
\end{theorem}

With $\calT \subseteq \calF$ the next corollary follows immediately.

\begin{corollary}\label{cor:tree-stability-vertexstability}
  For all $G \in \calT$ and $\xi \in \{ \alpha, \beta, \omega,\chi \}$,
  the problems
  $\xi$-$\textsc{Stability}$ and
  $\xi$-$\textsc{VertexStability}$ belong to $\Pol$.
\end{corollary}

We now focus on the unfrozenness problems.  All trees and forests with
fewer than three vertices ($I_1$, $I_2$, and~$P_2$) were already
covered in previous sections.  It remains to study trees and forests
with at least three vertices.

\begin{proposition}\label{prop:tree-omega-chi-unfrozenness}
  Every tree $G \in \calT$ with $|V(G)| \geq 3$ is neither $\omega$-
  nor $\chi$-unfrozen.
\end{proposition}

Based on this result we can deduce whether forests are
$\omega$- or $\chi$-unfrozen. As forests without edges are
empty graphs, we study forests with at least one edge.

\begin{theorem}\label{thm:forest-omega-chi-unfrozenness}
  If $F \in \calF$ contains $P_2$ but no $P_3$ as
  induced subgraphs, $F$ is $\omega$- and $\chi$-unfrozen. If $F$
  contains $P_3$ as an induced subgraph, $F$ is
  not
  $\omega$-
  nor
  $\chi$-unfrozen.
\end{theorem}

$\alpha$- and $\beta$-$\unfrozenness$
are
covered
in \cref{cor:tree-forest-alpha-beta-unfrozenness} of the
next subsection.

\subsection{Bipartite Graphs}\label{subsec:bipartite-graphs}

$G = (V_1 \cup V_2, E)$ is a \emph{bipartite graph} if $V_1 \cap V_2 =
\emptyset$ and $E \subseteq V_1 \times V_2$.  Denote the set of all
bipartite graphs by~$\calB$. We begin with two simple observations.
Again,
most proofs are deferred to the appendix.

\begin{observation}\label{obs:bipartite-graph-chi-omega}
  Let $G \in \calB$ be a bipartite graph. Then
  $\chi(G) = \omega(G) = 1$ if $E(G) = \emptyset$, and
  $\chi(G) = \omega(G) = 2$ if $E(G) \neq \emptyset$.
\end{observation}

Consequently, we can efficiently calculate $\chi$ and $\omega$ for all
bipartite graphs.  Next, we describe a tractable method to
calculate $\alpha$ and $\beta$ for bipartite graphs.

\begin{observation}\label{obs:bipartite-alpha-beta-efficiently}
 We can calculate $\alpha(G)$ and $\beta(G)$ efficiently for $G \in \calB$.
\end{observation}

Hence, we can efficiently calculate $\xi(G)$ for every $G \in \calB$
and $\xi \in \{ \alpha, \beta, \omega, \chi \}$.  Furthermore, as the
class of bipartite graphs is closed under subgraphs and induced
subgraphs, the following corollary follows from
Theorem~\ref{thm:general-efficient-problems}.

\begin{corollary}\label{cor:bipartite-stability-vertexstability}
  For every $\xi \in \{ \alpha, \beta, \omega, \chi \}$, the problems
  $\xi$-$\stability$ and $\xi$-$\vstability$ are in $\Pol$ for all
  bipartite graphs.
\end{corollary}

Next, we discuss approaches for how to decide whether a bipartite
graph is stable.  If a bipartite graph $G$ has no edges, we have
$G = I_{|V(G)|}$.
For bipartite graphs with one edge, we have the following simple
result.

\begin{proposition}\label{prop:bipartite-one-edge}
  Let $G$ be a bipartite graph with $|E(G)| = 1$.  Then $G$ is neither
  $\xi$-stable nor $\xi$-vertex-stable for $\xi \in\{ \alpha, \beta,
  \omega, \chi \}$.
\end{proposition}

Next, we provide results for bipartite graphs with more than one edge.

\begin{lemma}\label{lem:bipartite-more-edges-chi}
  Every bipartite graph $G$ with $|E(G)| \geq 2$ is $\chi$-stable.
\end{lemma}

With Lemma~\ref{lem:bipartite-more-edges-chi} we can characterize
$\chi$-vertex-stability.

\begin{theorem}\label{thm:bipartite-chi-vertexstability-characterization}
  Let $G$ be a bipartite graph with $|E(G)| \geq 2$.  $G$ is
  $\chi$-vertex-stable if and only if for all $v \in V(G)$ it holds that
  $\text{deg}(v) < |E(G)|$.
\end{theorem}

The proof of the following lemma is similar to that of
Lemma~\ref{lem:bipartite-more-edges-chi}.

\begin{lemma}\label{lem:bipartite-more-edges-omega}
  Every bipartite graph $G$ with $|E(G)| \geq 2$ is $\omega$-stable.
\end{lemma}

With Lemma~\ref{lem:bipartite-more-edges-omega} we also can
characterize $\omega$-vertex-stability.

\begin{theorem}\label{thm:bipartite-omega-vertex-stable}
  Let $G$ be a bipartite graph with $|E(G)| \geq 2$.
  $G$ is $\omega$-vertex-stable if and only if for all $v \in
  V(G)$ it holds that $\text{deg}(v) < |E(G)|$.
\end{theorem}

Besides these (vertex-)stability characterizations for bipartite
graphs, we now address unfrozenness
for
them.

\begin{theorem}\label{thm:bipartite-chi-unfrozenness}
  Let $G$ be a bipartite graph.  $G$ is $\chi$-unfrozen if
  and only if $G$ possesses no $P_3$ as an induced subgraph.
\end{theorem}

\begin{proofs}
  We prove both directions separately.  First, assume $G$ is
  $\chi$-unfrozen but contains $P_3$ as an induced subgraph. Write
  $V(P_3) = \{ v_1, v_2, v_3 \}$ and $E(P_3) = \{ \{ v_1, v_2 \}, \{
  v_2, v_3 \} \}$ for the corresponding vertices and edges.
  Then $e = \{ v_1, v_3 \} \in \overline{E}(G)$.
  However, adding $e$ to $G$ we obtain $\chi(G) = 2 < 3 = \chi(G + e)$,
  as $P_3 + e$ forms a $3$-clique in~$G$, a contradiction to the
  assumption that $G$ is $\chi$-unfrozen.
  Next, assume that $G$ possesses no $P_3$ as
  an induced subgraph but is not $\chi$-unfrozen. Hence, there must
  exist $e = \{ u, v \} \in \overline{E}(G)$ such that $\chi(G + e) = 3
  > 2 = \chi(G)$. Denote the two disjoint vertex sets of $G$ by
  $V_1 \cup V_2 = V(G)$. Obviously, $u \in V_1$ and $v \in V_2$ cannot
  be true, since then $\chi(G + e) = 2$ would hold. Therefore, without
  loss of generality, we assume $u, v \in V_1$. Adding $e$ to $G$ must
  create a cycle of odd length in~$G$, as cycles of even length as well
  as paths can be colored with two colors. Consequently, $G + e$
  possesses $C_n$ with $n = 2k + 1 \geq 3$, $k \in \N$, as a subgraph.
  This implies that $G$ must possess $P_3$ as an induced subgraph, again a
  contradiction.
\end{proofs}

Slightly modifying (the direction from right to left in)
the previous proof yields
Corollary~\ref{cor:bipartite-omega-unfrozenness}.
This time, adding $e$ to $G$ must create a $3$-clique in~$G$.

\begin{corollary}\label{cor:bipartite-omega-unfrozenness}
  $G \in \calB$ is $\omega$-unfrozen if and only if
  $G$ possesses no $P_3$ as an induced subgraph.
\end{corollary}

 Both results show that $\omega$- and
$\chi$-$\unfrozenness$ belong to $\Pol$ for bipartite graphs. For the
last results of this section we require the following lemma.

\begin{lemma}\label{lem:bipartite-vertex-cover-with-u}
  Let $G \in \calB$ be a bipartite graph and $u \in V(G)$. If $\beta(G
  - u) = \beta(G) - 1$, then there exists some vertex cover $V'
  \subseteq V(G)$ with $u \in V'$ and $|V'| = \beta(G)$.
\end{lemma}

\begin{theorem}\label{thm:bipartite-beta-unfrozenness}
  For every $G \in \calB$, the problem
  $\beta$-$\unfrozenness$ belongs to~$\Pol$.
\end{theorem}

The previous proof
allows
for every nonedge of a bipartite
graph
to decide if it is $\beta$-unfrozen
such that $\beta$-$\frozenness \in \Pol$ follows for
$G \in \calB$.
Gallai's theorem~\cite{gal:j:extreme-punkt-kantenmengen}
immediately yields
$\alpha$-$\unfrozenness \in \Pol$
for bipartite graphs.

\begin{corollary}\label{cor:bipartite-alpha-unfrozenness}
  $\alpha$-$\unfrozenness$ and $\beta$-$\frozenness \in \Pol$
  for all $G \in \calB$.
\end{corollary}

Since $\calT \subseteq \calF \subseteq \calB$, the next
corollary follows as well.

\begin{corollary}\label{cor:tree-forest-alpha-beta-unfrozenness}
  The problems $\alpha$- and $\beta$-$\unfrozenness$ as well as the
  problem $\beta$-$\textsc{Frozenness}$ belong to $\Pol$ for all trees
  and forests.
\end{corollary}

\subsection{Co-Graphs}\label{subsec:co-graphs}

First of all, we recursively define co-graphs, following a slightly
adjusted definition by Corneil
et al.~\cite{cor-ler-bur:j:complement-reducible-graphs}.

\begin{definition}[co-graph]
  The graph $G = (\{ v \}, \emptyset)$ is a \emph{co-graph}. If $G_1$
  and $G_2$ are co-graphs, then $G_1 \cup G_2$ and $G_1 + G_2$ are
  co-graphs, too.
\end{definition}

We denote the set of all co-graphs by $\calC$ and use the operators $\cup$ and $+$ as is common
(see, e.g., \cite{fre-hem-rot:c:complexity-of-stability}).
We will use the following result by
Corneil et al.~\cite{cor-ler-bur:j:complement-reducible-graphs}.

\begin{theorem}\label{thm:cograph-properties}
  Co-graphs are (i) closed under complements and (ii) closed under
  induced subgraphs,  but (iii) not closed under subgraphs in
  general. Furthermore, $G \in \calG$ is a co-graph if and only if $G$
  possesses no $P_4$ as an induced subgraph.
\end{theorem}

Property~(iii) is not proven in
their work~\cite{cor-ler-bur:j:complement-reducible-graphs}.
However, $C_4\in \calC$ is an easy example since $C_4$ is a
co-graph (see \cref{ex:co-graph-c4} below), and removing one edge yields
$P_4$.
Since every co-graph can be
constructed by
$\cup$ and $+$, we can identify a
co-graph by its \emph{co-expression}.

\begin{example}[co-expression]\label{ex:co-graph-c4}
  The co-expression $X = (v_1 \cup v_3) + (v_2 \cup v_4)$ describes the
  graph $C_4 = (\{ v_1, v_2, v_3, v_4 \}, \{ \{ v_1, v_2 \}, \{ v_2,
  v_3 \}, \{ v_3, v_4 \}, \{ v_4, v_1 \}\})$.
\end{example}

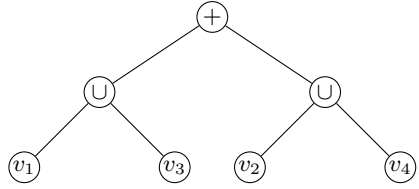
\begin{figure}[t!]
  \centering
  \begin{tikzpicture}
  \def \gnode {node[circle, draw,
    inner sep=0pt, minimum width=3ex]};
  \draw (0,0) \gnode (a) {$+$};
  \draw (-1.5,-1) \gnode (b) {$\cup$};
  \draw (1.5,-1) \gnode (c) {$\cup$};
  \draw (-2.5,-2) \gnode (v1) {$v_1$};
  \draw (-0.5,-2) \gnode (v3) {$v_3$};
  \draw (0.5,-2) \gnode (v2) {$v_2$};
  \draw (2.5,-2) \gnode (v4) {$v_4$};
  \draw[-] (a) to (b);
  \draw[-] (a) to (c);
  \draw[-] (b) to (v1);
  \draw[-] (b) to (v3);
  \draw[-] (c) to (v2);
  \draw[-] (c) to (v4);
\end{tikzpicture}
  \caption{Co-tree for $C_4$.}
  \label{fig:co-tree-c4}
\end{figure}
Obviously, we can build a binary tree for every co-graph via its
co-expression. The tree's leaves correspond to the graph's vertices
and the inner nodes of the tree correspond to the expression's
operations. For example, the tree in \cref{fig:co-tree-c4} corresponds
to the co-expression from \cref{ex:co-graph-c4} and, thus, describes
a~$C_4$.  We call such a tree a \emph{co-tree}.
To formulate our results regarding
stability and
unfrozenness of co-graphs, we need the following result of
Corneil
et al.~\cite{cor-per-ste:j:linear-recognition-algorithm-cographs}.

\begin{theorem}\label{thm:cograph-identify-tree}
  For every graph $G \in \calG$, we can decide in $\calO(|V(G)| +
  |E(G)|)$ time whether $G$ is a co-graph and, if so, provide a
  corresponding co-tree.
\end{theorem}

Combining the previous results with the next one by
Corneil et al.~\cite{cor-ler-bur:j:complement-reducible-graphs},
we can efficiently determine a co-graph's chromatic number.

\begin{theorem}\label{thm:cograph-chromatic-number}
  Let $G \in \calG$ be a co-graph and $T$ the corresponding
  co-tree. For a node $w$ from~$T$, denote by $G[w]$ the
  graph induced by the subtree of $T$ with root~$w$. To every leave $v$
  of $T$ we add as a
  label $\chi(G[v]) = 1$. For every inner node $w$ of
  $T$ we add, depending on the inner node's type, the following label:
  (1)~If $w$ is a $\cup$-node with children $v_1$ and $v_2$,
    $\chi(G[w]) = \max \{ \chi(G[v_1]), \chi(G[v_2]) \}$, and
    (2)~if $w$ is a $+$-node with children $v_1$ and $v_2$,
    $\chi(G[w]) = \chi(G[v_1]) + \chi(G[v_2])$.
  If $r$ is the root node of $T$, then it holds that $\chi(G[r]) =
  \chi(G)$.
\end{theorem}

A result similar to the previous one for $\alpha$ was given by Corneil
et al.~\cite{cor-ler-bur:j:complement-reducible-graphs}.

\begin{remark}\label{rem:co-graph-independent-set}
  We label all leaves of $T$ with $\alpha(G[v]) = 1$.
  Each inner node $w$ of $T$ with children $v_1$ and $v_2$ is labeled
  by $\alpha(G[w]) = \max\{ \alpha(G[v_1]), \alpha(G[v_2]) \}$ if $w$
  contains the
  $+$-operation, and by $\alpha(G[w]) = \alpha(G[v_1]) + \alpha(G[v_2])$ if
  $w$ contains the $\cup$-operation. For the root $r$ of~$T$,
  it then holds that $\alpha(G[r]) = \alpha(G)$.
\end{remark}

By the previous remark we can efficiently calculate $\alpha$ for
co-graphs. Based on these results, we can state the following theorems
whose proofs again are deferred to the appendix.

\begin{theorem}\label{thm:co-graph-chi-vertex-stability}
  For every $G \in \calC$, the problem
  $\chi$-$\textsc{VertexStability}$ is in~$\Pol$.
\end{theorem}

With a  similar proof as for the previous theorem, we obtain
the next result.

\begin{theorem}\label{thm:cograph-alpha-vertexstability}
  For every $G \in \calC$, the problem
  $\alpha$-$\textsc{VertexStability}$ is in $\Pol$.
\end{theorem}

We can use the same proof as for
\cref{thm:cograph-alpha-vertexstability} to obtain the next
corollary. However, this time we additionally use Gallai's
theorem~\cite{gal:j:extreme-punkt-kantenmengen} to calculate $\beta$ out of
$\alpha$ for $G$ and all induced subgraphs with one vertex removed.

\begin{corollary}\label{cor:co-graph-beta-vertex-stability}
  For every co-graph, the problem $\beta$-$\textsc{VertexStability}$ is
  in $\Pol$.
\end{corollary}

Although Frei et
al.~\cite[Proposition~5(5)]{fre-hem-rot:c:complexity-of-stability}
have already shown that the problem $\omega$-$\vstability$ is in
$\Pol$ for co-graphs, the next corollary provides an alternative
because $\alpha$-$\vstability = \{ \overline{G} \mid G \in
\omega\text{-}\vstability \}$ is true and co-graphs are closed under
complements.

\begin{corollary}\label{cor:co-graph-omega-vertex-stability}
  For all $G \in \calC$ the problem $\omega$-$\vstability$ is
  in $\Pol$.
\end{corollary}

Next, let us study the edge-related stability problems for
co-graphs. To obtain our results, we need the following two
auxiliary propositions.

\begin{proposition}\label{prop:cograph-critical-vertex-subgraph}
  Let $G \in \calC$ with $|V(G)| > 1$ and let $u \in V(G)$ be
  $\chi$-critical for~$G$.  There exist two co-graphs
  $G_1, G_2$ such that $G = G_1 \cup G_2$ or $G = G_1 + G_2$.
   Assuming, without loss of generality, that $u \in V(G_1)$, $u$ is
   $\chi$-critical for~$G_1$.
\end{proposition}

\begin{proposition}\label{prop:co-graph-critical-vertices-critical-edge}
  Let $G \in \calC$ and $e = \{ u, v \} \in E(G)$. If $u$ and $v$ are
  $\chi$-critical for~$G$, then $e$ is $\chi$-critical for $G$ as well.
\end{proposition}

Having these results, we are now able to provide our
stability-related results.

\begin{theorem}\label{thm:co-graph-chi-stability}
  For all co-graphs, the problem $\chi$-$\textsc{Stability}$ is
  in $\Pol$.
\end{theorem}

Next, we want to study the problem of $\omega$-$\textsc{Stability}$
for co-graphs. To do so, we need the following lemmas with their proofs
deferred to the appendix.

\begin{lemma}\label{lem:co-graph:set-of-cliques}
  Let $G \in \calC$ be a co-graph. We can compute \emph{all}
  cliques of size $\omega(G)$ for $G$ in time polynomial in~$|G|$.
\end{lemma}

\begin{lemma}\label{lem:co-graph-omega-reduce-by-one}
  Let $G \in \calG$ be a graph and $v \in V(G)$ and $e \in E(G)$.
  Then it holds that $\omega(G - v)$ and $\omega(G - e)$ are in
  $\{ \omega(G) - 1, \omega(G) \}$.
\end{lemma}

Having these results, we can show the next theorem.

\begin{theorem}\label{thm:co-graph-omega-stability}
  The problem $\omega$-$\textsc{Stability}$ is in $\Pol$ for co-graphs.
\end{theorem}

As we now know that we can efficiently determine whether a given
co-graph $G$ is $\omega$-stable, we can exploit the fact that
co-graphs are closed under complements to obtain the following
corollary (whose proof is deferred to the appendix).

\begin{corollary}\label{cor:co-graph-alpha-stability}
  The problem $\alpha$-$\textsc{Stability}$ is in $\Pol$ for co-graphs.
\end{corollary}

The next result follows
from Gallai's theorem~\cite{gal:j:extreme-punkt-kantenmengen}
and
\cite[Proposition~5]{fre-hem-rot:c:complexity-of-stability}.

\begin{corollary}\label{cor:co-graph-beta-stability}
  The problem $\beta$-$\textsc{Stability}$ is in $\Pol$ for co-graphs.
\end{corollary}

At this point, we finish the study of stability problems for
co-graphs, as all open questions are answered, and turn to the
problems related to unfrozenness.  The next two corollaries exploit
the fact that co-graphs are closed under complements and follow a
similar argumentation.

\begin{corollary}\label{cor:co-graph-alpha-beta-unfrozenness}
  The problems $\beta$-$\unfrozenness$ and $\alpha$-$\unfrozenness$
  are in $\Pol$ for co-graphs.
\end{corollary}

\begin{corollary}\label{cor:co-graph-omega-unfrozenness}
  The problem $\omega$-$\unfrozenness$ is in $\Pol$ for co-graphs.
\end{corollary}

Finally, we answer the last remaining open question related to
unfrozenness and co-graphs.

\begin{theorem}\label{thm:co-graph-chi-unfrozenness}
  The problem $\chi$-$\unfrozenness$ is in $\Pol$ for co-graphs.
\end{theorem}

\section{Conclusion}
\label{sec:conclusion}

We have provided 84 tractability results regarding the stability,
vertex-stability, and unfrozenness problems when restricted to special
graph classes. In particular, we studied these three problems for
seven important graph classes and four central graph parameters. Doing
so, our work provides some baseline for further, more expanding work
along this line of research.  For future work, we propose to study
further special graph classes that are not covered here.
Besides the study of stability for other graph classes, one can
also study the concept of \emph{cost of stability}:\footnote{Bachrach
  et
  al.~\cite{bac-elk-mal-mei-pas-ros-rot-zuc:j:bounds-on-the-cost-of-stabilizing-a-cooperative-game}
  study a related notion of ``cost of stability'' for cooperative
  games.}
Given a graph, the question is how costly it is to stabilize it.  In
other words, what is the smallest number of vertices or edges to be
added to or removed from the graph such that the resulting graph is
stable or unfrozen with respect to some graph parameter.  Relatedly,
it would make sense to combine these two approaches and study the
cost of stability for special graph classes.

\section*{Acknowledgments}
This work was supported in part by Deutsche Forschungsgemeinschaft under
grants RO~1202/14-2 and RO~1202/21-1.
\bibliographystyle{splncs04}
\bibliography{sources.bib}

\appendix

\section{Deferred Proofs from Section~\ref{sec:general-stability}:
  General Stability and Unfrozenness Results}

\sproofof{Theorem~\ref{thm:general-closed-subgraph-stability-in-p}}
  Let $G \in \calJ$ be a graph and $\xi$ a graph parameter that can be
  computed efficiently for graphs in $\calJ$. Now, compute
  $\xi(G)$. Since $\calJ$ is closed under subgraphs, for every every
  edge $e \in E(G)$ it holds that $G - e$ is in $\calJ$. Hence, we can
  compute $\xi(G - e)$ efficiently. Simply check for every edge whether
  $\xi(G - e) \neq \xi(G)$ holds. If no such edge exists, we know that
  $G$ is $\xi$-stable. Consequently, we can solve
  $\xi$-$\textsc{Stability}$ for all graphs in $\calJ$
  in~$\Pol$.~\eproofof{Theorem~\ref{thm:general-closed-subgraph-stability-in-p}}

\sproofof{Theorem~\ref{thm:perfect}}
  Let $G$ be a perfect graph. We have $\chi(G) =
  \omega(G)$ as well as for every vertex $v \in V(G)$ it holds that $\chi(G -
  v) = \omega(G - v)$ as $G - v$ is an induced subgraph of~$G$.
  Consequently, we have
  \begin{align*}
    G \in \omega\text{-}\vstability & \Leftrightarrow \forall{v \in
    V(G)} \colon \omega(G - v) = \omega(G) \\ & \Leftrightarrow \forall{v
    \in V(G)} \colon \chi(G - v) = \chi(G) \\ & \Leftrightarrow G\ \in
    \chi\text{-}\vstability.
  \end{align*}
  This completes the proof.~\eproofof{Theorem~\ref{thm:perfect}}

\sproofof{Theorem~\ref{thm:general-closed-complement-subgraph-omega-unfrozenness-in-p}}
  Without loss of generality, assume that we can efficiently compute
  $\alpha(G)$ for all $G \in \calJ$. Since $\calJ$ is closed under
  subgraphs, applying \cref{thm:general-closed-subgraph-stability-in-p}
  yields that $\alpha$-$\stability$ belongs to $\Pol$ for all graphs in
  $\calJ$. Let $G \in \calJ$ be a graph. As $\calJ$ is closed under
  complements, $\overline{G}$ also belongs to $\calJ$. Consequently, we
  can efficiently compute $\alpha(\overline{G})$. From Proposition 5
  1. in \cite{fre-hem-rot:c:complexity-of-stability} we know that
  $\alpha$-$\stability = \{ \overline{G} \mid G \in
  \omega$-$\unfrozenness \}$ holds. Hence, if $\overline{G}$ is
  $\alpha$-stable, it immediately follows that $G$ is
  $\omega$-unfrozen. Therefore, for graphs in $\calJ$ we can decide
  efficiently whether they are $\omega$-unfrozen, such that
  $\omega$-$\unfrozenness$ belongs to $\Pol$ for all graphs
  in~$\calJ$.~\eproofof{Theorem~\ref{thm:general-closed-complement-subgraph-omega-unfrozenness-in-p}}

\sproofof{Theorem~\ref{thm:general-other}}
  The argumentation is similar to the argumentation of
  \cref{thm:general-closed-complement-subgraph-omega-unfrozenness-in-p}.
  We know that $\calJ$ is closed under subgraphs and we can
  compute $\omega(G)$ for all $G \in \calJ$ efficiently. Consequently,
  by \cref{thm:general-closed-subgraph-stability-in-p} it follows that
  $\omega$-$\stability$ belongs to $\Pol$ for all graphs in
  $\calJ$. From Proposition 5 2. in
  \cite{fre-hem-rot:c:complexity-of-stability} we know that
  \begin{align*}
    \beta\text{-}\unfrozenness = \alpha\text{-}\unfrozenness = \{
    \overline{G} \mid G \in \omega\text{-}\stability \}
  \end{align*}
  holds. Consequently, let $G \in \calJ$ be a graph. As $\calJ$ is
  closed under complements, $\overline{G}$ is in $\calJ$, too. Now,
  efficiently decide whether $\overline{G}$ is $\omega$-stable. If
  that is the case, it immediately follows that $G$ is $\alpha$- and
  $\beta$-unfrozen. Hence, $\alpha$- and $\beta$-$\unfrozenness$
  belong to $\Pol$ for all graphs
  in~$\calJ$.~\eproofof{Theorem~\ref{thm:general-other}}

\sproofof{Observation~\ref{obs:vertex-stability-subset-stability}}
  Let $G \in \chi$-$\textsc{VertexStability}$ be a graph. Consequently, $G$
  is $\chi$-vertex-stable, i.e., every $v \in V(G)$ is
  $\chi$-stable. Together with Observation~3
  in~\cite{fre-hem-rot:c:complexity-of-stability} it follows immediately
  that every edge $e \in E(G)$ is $\chi$-stable, since all its incident
  vertices are $\chi$-stable. If all edges of $G$ are $\chi$-stable, it
  follows that $G$ is $\chi$-stable and thus $G \in \chi$-$\textsc{Stability}$
  holds.~\eproofof{Observation~\ref{obs:vertex-stability-subset-stability}}

\sproofof{Observation~\ref{beta-critical-edge-critical-vertices}}
  This is basically the same proof as for Observation~2 due to Frei et
  al.~\cite{fre-hem-rot:c:complexity-of-stability}.  If $e$ is
  $\beta$-critical for $G$, we have $\beta(G - e) < \beta(G)$.
  Since $G - v$ and $G - u$ are induced subgraphs of $G - e$ it follows
  that $\beta(G - v), \beta(G - u) \leq \beta(G - e) < \beta(G)$ holds.
  Thus $u$ and $v$ are $\beta$-critical.~\eproofof{Observation~\ref{beta-critical-edge-critical-vertices}}

\section{Deferred Proofs from Section~\ref{subsec:empty-graphs}:
  Empty Graphs}

  \sproofof{Theorem~\ref{thm:empty-graphs-stability-vertexstability}}
  We prove every item on its own.
  \begin{enumerate}
  \item We have $\chi(I_0) = 0$ and $\chi(I_n) = 1$ for $n \geq 1$.
    Furthermore, for $v \in V(I_n)$ it holds that $I_n - v = I_{n-1}$,
    such that $\chi(I_n - v) = \chi(I_{n-1})$ holds.  Consequently, $I_n$
    for $n = 0$ or $n \geq 2$ is $\chi$-vertex-stable, but not for $n = 1$
    since the only vertex is $\chi$-critical.
  \item It holds that $\omega(I_0) = 0$ and $\omega(I_n) = 1$ for $n
    \geq 1$. Hence, $I_0$ and $I_n$ for $n \geq 2$ are
    $\omega$-vertex-stable but $I_1$ isn't, since its only vertex is
    $\omega$-critical.
  \item We have $\beta(I_n) = 0$. Therefore, for every $v \in V(I_n)$
    we have $\beta(I_n) = \beta(I_n - v) = \beta(I_{n-1})$, so that all
    vertices are
    $\beta$-stable and, thus, $I_n$ is $\beta$-vertex-stable.
  \item It holds that $\alpha(I_n) = n$. Hence, for every $v \in
    V(I_n)$ we have $\alpha(I_n - v) = \alpha(I_{n-1}) = n - 1 < n =
    \alpha(I_n)$, so $I_n$ is not $\alpha$-vertex-stable for $n \geq 1$.
  \item Since $I_n$ for $n \geq 0$ has no edges, all edges of $I_n$
    are $\xi$-stable and, thus, $I_n$ is $\xi$-stable.
  \end{enumerate}
  This completes the
  proof.~\eproofof{Theorem~\ref{thm:empty-graphs-stability-vertexstability}}

  \sproofof{Proposition~\ref{prop:empty-graphs-chi-unfrozenness}}
  Since $\overline{E}(I_0) = \overline{E}(I_1) = \emptyset$, it
  obviously follows that both are $\chi$-unfrozen. For all other $I_n
  \in \calI$ with $n \geq 2$, we just add one arbitrary nonedge $e$ to
  $I_n$ so that $\chi(I_n + e) = 2 > 1 = \chi(I_n)$ holds. Consequently,
  these graphs cannot be $\chi$-unfrozen.~\eproofof{Proposition~\ref{prop:empty-graphs-chi-unfrozenness}}

  \sproofof{Corollary~\ref{cor:empty-graphs-unfrozenness}}
  It is trivial to see that $\overline{I_n} = K_n$ for all $n \in
  \N$. Consequently, by Proposition 5 2. from
  \cite{fre-hem-rot:c:complexity-of-stability} it follows that
  $\beta$-$\unfrozenness = \alpha$-$\unfrozenness$ belong to $\Pol$ for
  empty graphs, since $\omega$-$\stability$ is in $\Pol$ for all
  complete graphs, c.f. \cref{prop:complete-graphs-xi-critical}. With a
  similar argumentation it follows from Proposition 5 1. that
  $\omega$-$\unfrozenness$ is in $\Pol$ for all empty graphs, since
  $\beta$-$\stability$ is in $\Pol$ for all complete graphs.~\eproofof{Corollary~\ref{cor:empty-graphs-unfrozenness}}

\section{Deferred Proofs from
  Section~\ref{subsec:complete-graphs}: Complete Graphs}

  \sproofof{Observation~\ref{obs:complete-graphs-chi-vertex-criticality}}
  Let $K_n \in \calK$ be a complete graph. Obviously, for every vertex
  $v \in V(K_n)$ we have $K_n - v = K_{n-1}$ as well as
  $\chi(K_n) = n$. Consequently, for every $v \in V(K_n)$ we have
  \begin{align*}
    \chi(K_n - v) = \chi(K_{n-1}) = n - 1 < n = \chi(K_n),
  \end{align*}
  such that $K_n$ is
  $\chi$-vertex-critical.~\eproofof{Observation~\ref{obs:complete-graphs-chi-vertex-criticality}}

  \sproofof{Proposition~\ref{prop:complete-graphs-xi-critical}}
  For every $e \in E(K_n)$ we have $\omega(K_n - e) = n - 1 < n =
  \omega(K_n)$, such that $e$ is $\omega$-critical. Furthermore, we have
  $\chi(K_n - e) = n - 1 < n = \chi(K_n)$, $\alpha(K_n - e) = 2 > 1 =
  \alpha(K_n)$ and $\beta(K_n - e) = n - 2 < n - 1 = \beta(K_n)$, so
  that $e$ is $\alpha$-, $\chi$- and $\beta$-critical. The previous
  argumentation holds for all edges in $E(K_n)$, and therefore, $K_n$ is
  $\xi$-critical.
  \eproofof{Proposition~\ref{prop:complete-graphs-xi-critical}}

\section{Deferred Proofs from Section~\ref{subsec:paths}: Paths}

  \sproofof{Observation~\ref{obs:path-parameters}}
  Obviously, $\chi(P_0) = 0$ and $\chi(P_1) = 1$ are true. Furthermore,
  for $n \geq 2$ we simply color the vertices of $P_n$ alternatingly in
  two colors. Additionally, one can immediately see $\omega(P_0) = 0$,
  $\omega(P_1) = 1$ as well as $\omega(P_n) = 2$ for $n \geq 2$.
  For the last statements, we argue as follows.  The graph $P_n$ has
  $n$ vertices and $n-1$ edges.  The first and the last vertex of the
  graph can each cover at most one edge.  All vertices in between can
  cover at most two edges.  Consequently, we need
    $\beta(P_n) = \left\lceil \frac{n-1}{2} \right\rceil = \left\lfloor
    \frac{n}{2} \right\rfloor$
  vertices for a vertex cover of~$P_n$.  Applying Gallai's
  theorem~\cite{gal:j:extreme-punkt-kantenmengen} yields $\alpha(P_n)
  = \lceil \frac{n}{2} \rceil$ from the previous result, as $|V(P_n)|
  = \beta(P_n) + \alpha(P_n)$ must hold.~\eproofof{Observation~\ref{obs:path-parameters}}

  \sproofof{Theorem~\ref{thm:path-stability-vertexstability}}
  \begin{enumerate}
  \item $P_1$ has no edges, so $P_1$ is $\xi$-stable.  Furthermore,
    since $P_1 - v_1 = P_0$ and $\beta(P_1) = \beta(P_0) = 0$ holds,
    $P_1$ is $\beta$-vertex-stable, too.  Obviously, $P_1$ is not
    $\chi$-vertex-stable, as $0 = \chi(P_0) = \chi(P_1 - v_1) <
    \chi(P_1) = 1$.  The same argumentation holds with respect to
    $\alpha$ and $\omega$.
  \item We have $\chi(P_2) = 2, \beta(P_2) = 1, \alpha(P_2) = 1,
    \omega(P_2) = 2$ as well as $\chi(P_2 - \{ v_1, v_2 \}) = 1,
    \beta(P_2 - \{ v_1, v_2 \}) = 0, \alpha(P_2 - \{ v_1, v_2 \}) = 2,
    \omega(P_2 - \{ v_1, v_2 \}) = 1$, so $P_2$ is not $\xi$-stable.
    Furthermore, we have $\chi(P_2 - v_1) = 1, \beta(P_2 - v_1) = 0$
    and $\omega(P_2 - v_1) = 1$, so $P_2$ is neither
    $\rho$-vertex-stable.  However, $\alpha(P_2 - v_1) = \alpha(P_2 -
    v_2) = 1$, so $P_2$ is $\alpha$-vertex-stable.
  \item We have $\chi(P_3) = 2, \beta(P_3) = 1, \alpha(P_3) = 2$ and
    $\omega(P_3) = 2$. For all $e \in E(P_3)$ we have $\chi(P_3 - e)
    = 2$, $\beta(P_3 - e) = 1, \alpha(P_3 - e) = 2$ and $\omega(P_3
    - e) = 2$, as $E(P_3 - e) \neq \emptyset$, i.e., $P_3$ is
    $\xi$-stable. Since we have $\chi(P_3 - v_2) = 1, \beta(P_3 -
    v_2) = 0, \alpha(P_3 - v_1) = 1$ and $\omega(P_3 - v_2) = 1$, it
    follows that $P_3$ is not $\xi$-vertex-stable.
  \item Let $n \geq 4$. For all $v \in V(P_n)$ we have $\chi(P_n -
    v) = \omega(P_n - v) = 2$, as $E(P_n - v) \neq
    \emptyset$. Consequently, $P_n$ is $\chi$- and
    $\omega$-vertex-stable. The same argument holds for all $e \in
    E(P_n)$, so $P_n$ is $\chi$- and $\omega$-stable, too. From
    \cref{obs:path-parameters} we know that $\beta(P_n) = \lfloor
    \frac{n}{2} \rfloor$ holds. Consequently, we have
    \begin{align*}
      \beta(P_n - v_{n-1}) = \beta(P_{n-2} \cup P_1) =
      \beta(P_{n-2}) = \left\lfloor \frac{n-2}{2} \right\rfloor
      < \left\lfloor \frac{n}{2}
      \right\rfloor = \beta(P_n),
    \end{align*}
    so $P_n$ is not $\beta$-vertex-stable.

    If $n = 2k$ for a $k \in \N$, we have $\beta(P_n) = k$ as well
    as
    \begin{align*}
      \beta(P_n - \{ v_n, v_{n-1} \}) = \beta(P_{2k - 1} \cup P_1) =
      \beta(P_{2k-1}) = k - 1,
    \end{align*}
    so $P_n$ is not $\beta$-stable. Furthermore,
    it holds that $\alpha(P_n) = k$ and
    \begin{align*}
      \alpha(P_n - \{ v_{n-1}, v_n \}) = \alpha(P_{n-1} \cup P_1) =
      k + 1,
    \end{align*}
    so $P_n$ is neither $\alpha$-stable.  However, let $v \in
    V(P_n)$.  Consequently, we have
    \begin{align*}
      \alpha(P_n - v) = \alpha(P_p \cup P_q)
    \end{align*}
    with $ p + q = n - 1$, assuming w.l.o.g. $p = 2s + 1, q = 2t, s,
    t \in \N$. Then it follows that
    \begin{align*}
      \alpha(P_p \cup P_q) = \alpha(P_p) + \alpha(P_q) = s + 1 + t = k,
    \end{align*}
    so $P_n$ is $\alpha$-vertex-stable. Next, assume $n = 2k
    + 1$. We have $\beta(P_n)= k$ and for every $e \in E(P_n)$ it
    holds that
    \begin{align*}
      \beta(P_{n} - e) = \beta(P_p \cup P_q)
    \end{align*}
    for $p,q \in \N$ with $p + q = n$. Without loss of
    generality we can assume that $p = 2s$ and $ q = 2t + 1$ for
    $s,t \in \N$. Then it holds that
    \begin{align*}
      \beta(P_p \cup P_q) = \beta(P_p) + \beta(P_q) = \left\lfloor
      \frac{p}{2} \right\rfloor + \left\lfloor \frac{q}{2} \right\rfloor = s
      + t = \frac{2(s+t)}{2} = \frac{n - 1}{2} = k.
    \end{align*}
    Therefore, $\beta(P_n - e) = k = \beta(P_n)$ and, thus, $P_n$ is
    $\beta$-stable. Additionally, it holds that $\alpha(P_n) = k +
    1$ and for every $e \in E(P_n)$ we have
    \begin{align*}
      \alpha(P_n - e) = \alpha(P_p \cup P_q)
    \end{align*}
    with $p + q = n$, assuming without loss of generality that $p =
    2s$ and $q = 2t + 1$ for $s, t, \in \N$. Then it follows that
    \begin{align*}
      \alpha(P_p \cup P_q) = \alpha(P_p) + \alpha(P_q) = s + t + 1 =
      k +1.
    \end{align*}
    Therefore, $P_n$ is $\alpha$-stable. Finally, it holds that
    \begin{align*}
      \alpha(P_n - v_n) = \alpha(P_{n-1}) = k,
    \end{align*}
    so $P_n$ is not $\alpha$-vertex-stable.~\eproofof{Theorem~\ref{thm:path-stability-vertexstability}}
  \end{enumerate}

  \sproofof{Observation~\ref{obs:path-2-3-unfrozen-frozen}}
  The first observation related to $P_2$ immediately follows from the
  fact that $\overline{E}(P_2) = \emptyset$ holds, because then every
  nonedge of $P_2$ is $\xi$-unfrozen and hence $P_2$ is
  $\xi$-unfrozen, too. For the second observation one must note that
  $\overline{E}(P_3) = \{\{ v_1, v_3 \}\}$ holds, i.e., there is only
  one nonedge. When we add this nonedge to $P_3$ we obtain
  $K_3$. As $\chi(P_3) = 2 < 3 = \chi(K_3)$, $\alpha(P_3) = 2 > 1 =
  \alpha(K_3)$, $\beta(P_3) = 1 < 2 = \beta(K_3)$ and $\omega(P_3) = 2
  < 3 = \omega(K_3)$ hold, it follows that $\{ v_1, v_3 \}$ is
  $\xi$-frozen, such that $P_3$ is $\xi$-frozen, too.~\eproofof{Observation~\ref{obs:path-2-3-unfrozen-frozen}}

  \sproofof{Proposition~\ref{prop:path-omega-chi-unfrozen}}
  From \cref{obs:path-parameters} we know $\chi(P_n) = \omega(P_n) =
  2$. When we add the nonedge $e = \{ v_1, v_3 \} \in
  \overline{E}(P_n)$ to $P_n$, we obtain $\chi(P_n + e) = 3$ as well as
  $\omega(P_n + e) = 3$, since $v_1, v_2$ and $v_3$ then build a
  $3$-clique. Thus $e$ is $\chi$- and $\omega$-frozen such that $P_n$
  can neither be $\chi$- nor $\omega$-unfrozen.~\eproofof{Proposition~\ref{prop:path-omega-chi-unfrozen}}

  \sproofof{Proposition~\ref{prop:path-alpha-beta-unfrozenness}}
  For this result we require two facts. For $n \in \N$ denote
  by $C_n \in \calG$ a circle with $n$ vertices. Then we have
  $\beta(C_n) = \lceil \frac{n}{2} \rceil$ as well as $\alpha(C_n) =
  \lfloor \frac{n}{2} \rfloor$. Furthermore, a path $P_n$ with $n = 2k
  \in \N$ vertices has two vertex covers of size $\beta(P_n)$, namely
  $B_1 = \{ v_{2i} \mid 1 \leq i \leq k \}$ and $B_2 = \{ v_{2i+1} \mid
  0 \leq i < k \}$.
  We prove both statements separately. First, assume $n = 2k + 1$ for $k
  \in \N$. In this case $e = \{ v_1, v_n \} \in \overline{E}(P_n)$ is a
  nonedge for $P_n$. Adding $e$ to $P_n$ results in $P_n + e =
  C_n$. With the previous remark we know that
  \begin{align*}
    \beta(P_n) = \left\lfloor \frac{n}{2} \right\rfloor = k < k + 1 =
    \left\lceil \frac{n}{2} \right\rceil = \beta(C_n) = \beta(P_n + e),
  \end{align*}
  such that $e$ is $\beta$-frozen and, thus, $P_n$ cannot be
  $\beta$-unfrozen. A similar argument with respect to $\alpha$,
  \begin{align*}
    \alpha(P_n + e) = \alpha(C_n) = \left\lfloor \frac{n}{2}
    \right\rfloor = k < k + 1 = \left\lceil \frac{n}{2} \right\rceil =
    \alpha(P_n),
  \end{align*}
  yields that $e$ is $\alpha$-frozen and consequently, $P_n$ cannot be
  $\alpha$-unfrozen.

  Now, assume that $n = 2k$ for $k \in \N$ holds, i.e., $n$ is
  even. Denote by $e = \{ v_i, v_j \} \in \overline{E}(P_n)$ a
  nonedge of $P_n$, i.e., $j \neq i + 1$ for $1 \leq i < n$. As
  previously mentioned, there are two optimal vertex covers for $P_n$,
  $B_1$, containing all evenly numbered vertices, and $B_2$, containing
  all oddly numbered vertices. Selecting $B_j, j \in \{ 1, 2 \}$, such
  that $B_j \cap e \neq \emptyset$ holds, results in an optimal vertex
  cover for $P_n + e$, also covering the newly introduced edge
  $e$. Consequently, $\beta(P_n + e) = \beta(P_n)$, such that $e$ is
  $\beta$-unfrozen. Since this argument holds for an arbitrary
  nonedge $e$, it follows that all nonedges are $\beta$-unfrozen and
  therefore, $P_n$ is $\beta$-unfrozen, too. If $P_n$ is
  $\beta$-unfrozen, for every nonedge $e \in \overline{E}(P_n)$ we have
  \begin{align*}
    \alpha(P_n + e) = |V(P_n + e)| - \beta(P_n + e) = |V(P_n)| -
    \beta(P_n) = \alpha(P_n),
  \end{align*}
  such that $P_n$ is $\alpha$-unfrozen,
  too.~\eproofof{Proposition~\ref{prop:path-alpha-beta-unfrozenness}}

\section{Deferred Proofs from
    Section~\ref{subsec:trees-forests}: Trees and Forests}

  \sproofof{Theorem~\ref{thm:forests-stability-vertexstability}}
  Let $\xi \in \{ \alpha, \beta, \omega, \chi \}$ and $G \in \calF$
  be a forest. As previously stated, we can efficiently calculate
  $\xi(G)$. Now, for every $v \in V(G)$ and $e \in E(G)$ it holds that
  $G - v, G - e \in \calF$, such that we can also efficiently compute
  $\xi(G - v)$ and $\xi(G - e)$. Consequently, we can decide in time
  polynomial in $|G|$ whether $G$ is $\xi$-vertex-stable or
  $\xi$-stable and therefore, both problems,
  $\xi$-$\textsc{Stability}$ and -$\textsc{VertexStability}$ belong to
  $\Pol$.~\eproofof{Theorem~\ref{thm:forests-stability-vertexstability}}

  \sproofof{Proposition~\ref{prop:tree-omega-chi-unfrozenness}}
  With $|V(G)| \geq 3$ and $|E(G)| = |V(G)| - 1$ it follows that $G$
  must contain $P_3$ as an induced subgraph. Denote the corresponding
  vertices by $v_1$, $v_2$, and $v_3$. Then $\{ v_1, v_2 \}, \{ v_2,
  v_3 \} \in E(G)$ is true. Furthermore, as $G$ does not contain
  any cycle, $e = \{ v_1, v_3 \}$ must be a nonedge of $G$. Adding $e$
  to $G$ creates the $3$-clique $v_1, v_2, v_3$ in $G + e$, such that
  we obtain
  \begin{align*}
    \omega(G) = 2 < 3 = \omega(G + e).
  \end{align*}
  Hence, $e$ is $\omega$-frozen and thus $G$ cannot be
  $\omega$-unfrozen. A similar argument yields that $G$ cannot be
  $\chi$-unfrozen.~\eproofof{Proposition~\ref{prop:tree-omega-chi-unfrozenness}}

  \sproofof{Theorem~\ref{thm:forest-omega-chi-unfrozenness}}
  We prove both statements separately. If $F$ contains $P_2$ but no
  $P_3$ as an induced subgraph, we have $\omega(F) = \chi(F) = 2$. Let $e =
  \{u,v\} \in \overline{E}(F)$ be a nonedge of $F$. Both vertices $u,v$
  satisfy one of two cases: Either the vertex is isolated or part of
  some $P_2$ in $F$. In both cases, adding $e$ to $F$ does not create a
  $3$-clique, such that $\omega(F + e) = \chi(F + e) = 2$ still holds
  and $e$ is $\omega$- and $\chi$-unfrozen. Since that holds for all
  nonedges of $F$, it follows that $F$ is $\omega$- and
  $\chi$-unfrozen. Contrarily, if $F$ contains $P_3$ as an induced
  subgraph, we can follow the same arguments as in the proof of
  \cref{prop:tree-omega-chi-unfrozenness} to see that $F$ is neither
  $\omega$- nor $\chi$-unfrozen.~\eproofof{Theorem~\ref{thm:forest-omega-chi-unfrozenness}}

\section{Deferred Proofs from
  Section~\ref{subsec:bipartite-graphs}: Bipartite Graphs}

  \sproofof{Observation~\ref{obs:bipartite-graph-chi-omega}}
  \begin{enumerate}
  \item If $E(G) = \emptyset$, it obviously holds that $\chi(G) = 1$,
    as we can color all vertices with the same color, and the largest
    clique has size $1$, such that $\omega(G) = 1$ is true, too.
  \item If $E(G) \neq \emptyset$, denote $V_1 \cup V_2 =
    V(G)$. Consequently, we can color all vertices in $V_1$ in one color
    and all vertices in $V_2$ in a second color, since there are no edges
    among the vertices of $V_1$ nor~$V_2$.
    Furthermore, there can not exist a clique of
    size three or larger in $G$, as bipartite graphs do not possess cycles
    of odd length as induced subgraphs, but every clique of size three or
    larger possesses a cycle of length three as an induced subgraph.~\eproofof{Observation~\ref{obs:bipartite-graph-chi-omega}}
  \end{enumerate}

  \sproofof{Observation~\ref{obs:bipartite-alpha-beta-efficiently}}
  Using the \emph{Hopcroft-Karp} algorithm~\cite{hop-kar:j:matching}
  we can efficiently compute a maximum matching $M \subseteq E(G)$ for
  $G$. Applying \emph{König's theorem}~\cite{koe:j:koenigs-theorem}, we
  know that $|M| = \beta(G)$ holds. Hence, $\beta(G)$ can be computed
  efficiently for every $G$. With Gallai's theorem we obtain $\alpha(G)$
  from $\beta(G)$ and, thus, can compute $\alpha(G)$ efficiently for $G$,
  too.~\eproofof{Observation~\ref{obs:bipartite-alpha-beta-efficiently}}

  \sproofof{Proposition~\ref{prop:bipartite-one-edge}}
  Denote $G$'s only edge by $e = \{ u, v \}$ for $u,v \in V(G)$. Then
  $G$ is neither $\alpha$-stable nor $\alpha$-vertex-stable, as
  $\alpha(G) = |V(G)| - 1$ as well as $\alpha(G - e) = |V(G)|$ and
  $\alpha(G - w) = |V(G)| - 2$ for $w \in V(G) \setminus \{ u, v \}$
  hold. Consequently, $G$ is neither $\beta$-stable, following from
  \cite[Proposition~5]{fre-hem-rot:c:complexity-of-stability}, nor
  $\beta$-vertex-stable because of $\beta(G) = 1$ and $\beta(G - u) =
  0$. Furthermore, we have $\omega(G) = 2$ as well as $\omega(G - e) =
  \omega(G - u) = 1$, such that $G$ is neither $\omega$-stable nor
  $\omega$-vertex-stable. Lastly, $\chi(G) = 2$ but $\chi(G - e) =
  \chi(G - u) = 1$, such that $G$ is not $\chi$-stable nor
  $\chi$-vertex-stable, too.~\eproofof{Proposition~\ref{prop:bipartite-one-edge}}

  \sproofof{Lemma~\ref{lem:bipartite-more-edges-chi}}
  Let $e \in E(G)$ be an arbitrary edge of~$G$. Since $E(G - e) \neq
  \emptyset$, it holds that $\chi(G - e) = 2 = \chi(G)$ and, thus, $G$
  is $\chi$-stable.~\eproofof{Lemma~\ref{lem:bipartite-more-edges-chi}}

  \sproofof{Theorem~\ref{thm:bipartite-chi-vertexstability-characterization}}
  Assume $G$ to be $\chi$-vertex-stable. Furthermore, as we assume
  that $|E(G)| \geq 2$, it holds that $\chi(G) = 2$. Then there cannot
  exist a vertex $v \in V(G)$ with $\text{deg}(v) = |E(G)|$, as such a
  vertex would be $\chi$-critical, since $\chi(G - v) = 1$ because of
  $E(G - v) = \emptyset$. For the opposite direction, assume that for
  all vertices $v \in V(G)$ we have $\text{deg}(v) < |E(G)|$. Hence, no
  matter what vertex $v \in V(G)$ we remove from $G$, it always holds
  that $E(G-v) \neq \emptyset$, so $\chi(G - v) = 2 = \chi(G)$ and,
  thus, $G$ is $\chi$-vertex-stable.~\eproofof{Theorem~\ref{thm:bipartite-chi-vertexstability-characterization}}

  \sproofof{Lemma~\ref{lem:bipartite-more-edges-omega}}
  For all $e \in E(G)$ we have $\omega(G - e) = 2 = \omega(G)$ as $E(G
  - e) \neq \emptyset$ holds, such that $G$ is $\omega$-stable.~\eproofof{Lemma~\ref{lem:bipartite-more-edges-omega}}

  \sproofof{Theorem~\ref{thm:bipartite-omega-vertex-stable}}
  Assume that $G$ is $\omega$-vertex-stable. Consequently, for all $v
  \in V(G)$ it holds that $\omega(G - v) = 2 = \omega(G)$. If there is
  one $v \in V(G)$ with $\text{deg}(v) = |E(G)|$, we have $\omega(G - v)
  = 1$ as $E(G - v) = \emptyset$, a contradiction to $G$'s
  $\omega$-vertex-stability. Contrarily, assume that for all $v \in
  V(G)$ it holds that $\text{deg}(v) < |E(G)|$. Then, for all $v \in
  V(G)$, it follows that $E(G - v) \neq \emptyset$. Consequently,
  $\omega(G) = 2 = \omega(G - v)$ and, hence, $G$ is
  $\omega$-vertex-stable.~\eproofof{Theorem~\ref{thm:bipartite-omega-vertex-stable}}

  \sproofof{Lemma~\ref{lem:bipartite-vertex-cover-with-u}}
  Denote by
  $V'' \subseteq V(G - u)$ a minimum vertex cover with $|V''| =
  \beta(G - u)$ and write
  $V' = V'' \cup \{ u \}$. Then $|V'| = \beta(G - u) + 1 =
  \beta(G)$
  and
  $E(G - u) \cup \{ e \in E(G) \mid e \cap \{ u \} \neq \emptyset \} =
  E(G)$ holds. As $V''$ is a
  minimum vertex cover for $G - u$, all edges in $E(G - u)$ are covered
  by $V''$. All remaining edges in $\{ e \in E(G) \mid e \cap \{ u \}
  \neq \emptyset \}$ are covered by $u$, so that $V'$ covers all edges
  in $E(G)$ and is a vertex cover of $G$.~\eproofof{Lemma~\ref{lem:bipartite-vertex-cover-with-u}}

  \sproofof{Theorem~\ref{thm:bipartite-beta-unfrozenness}}
  Let $G \in \calB$ be a bipartite graph with $V(G) = V_1 \cup V_2$
  and $V_1 \cap V_2 = \emptyset$. Then $E(G) \subseteq V_1 \times V_2$
  holds and, according to \cref{obs:bipartite-alpha-beta-efficiently},
  we can calculate $\beta(G)$ efficiently. For any nonedge $e = \{ u,v
  \} \in \overline{E}(G)$ either (1.) $e \in V_1 \times V_2$ or (2.) $e
  \in V_i \times V_i$, $i \in \{1,2\}$, must hold. We study both cases
  separately: (1) If $e \in V_1 \times V_2$, then $G + e$ is a bipartite
  graph, such that we can efficiently calculate $\beta(G + e)$ and
  compare it with $\beta(G)$ to determine whether $e$ is
  $\beta$-unfrozen or -frozen. (2) Without loss of generality assume $e
  \in V_1 \times V_1$. Then two cases are possible: (a) $G + e$ can be
  rearranged, such that it is bipartite. This is possible if and only if
  $\chi(G + e) = 2$, which can be checked efficiently. In this case we
  can compute $\beta(G + e)$ efficiently to determine whether $e$ is
  $\beta$-unfrozen or -frozen. (b) $G + e$ is no bipartite graph since
  it contains a cycle of odd length as subgraph. In this case we check
  with \cref{lem:bipartite-vertex-cover-with-u} for $u$, and afterwards
  for $v$, whether there exists some minimum vertex cover $V' \subseteq
  V(G)$ for $G$ with $u \in V'$ or $v \in V'$ respectively. If one of
  these two checks is positive, we know that $\beta(G + e) = \beta(G)$
  holds and hence, $e$ is $\beta$-unfrozen. Otherwise, $\beta(G + e) =
  \beta(G) + 1$ must hold, such that $e$ is $\beta$-frozen. Doing so, we
  can check every nonedge $e \in \overline{E}(G)$ efficiently for
  $\beta$-unfrozenness, such that $\beta$-$\unfrozenness$ is in $\Pol$
  for all graphs in $\calB$.~\eproofof{Theorem~\ref{thm:bipartite-beta-unfrozenness}}

\section{Deferred Proofs from Section~\ref{subsec:co-graphs}: Co-Graphs}

  \sproofof{Theorem~\ref{thm:cograph-alpha-vertexstability}}
  Let $G \in \calC$ be a co-graph. According to
  \cref{thm:cograph-identify-tree} we can calculate the graph's
  co-tree $T$ efficiently. Now, calculate $\chi(G)$ according to
  \cref{thm:cograph-chromatic-number}. Since co-graphs are closed
  under induced subgraphs, for every $v \in V(G)$ it holds that $G -
  v$ is a co-graph, too. Thus we can calculate $\chi(G - v)$
  efficiently. If there is a vertex $v \in V(G)$ such that $\chi(G -
  v) < \chi(G)$ holds, we immediately know that $G$ is not
  $\chi$-vertex-stable.  Otherwise, if for all $v \in V(G)$ it holds
  that $\chi(G - v) = \chi(G)$, it directly follows that $G$ is
  $\chi$-vertex-stable. Consequently, we can decide for every co-graph
  whether it is $\chi$-vertex-stable or not in $\Pol$ and, therefore,
  it follows that $\chi$-$\textsc{VertexStability}$ is in $\Pol$ for
  co-graphs.~\eproofof{Theorem~\ref{thm:cograph-alpha-vertexstability}}

  \sproofof{Theorem~\ref{thm:cograph-alpha-vertexstability}}
  Let $G \in \calC$ be a co-graph. Calculate $\alpha(G)$ according to
  \cref{rem:co-graph-independent-set}. Now, for every $v \in V(G)$ we
  calculate $\alpha(G - v)$ as previously described. If there exists at
  least one vertex $v \in V(G)$ such that $\alpha(G - v) \neq
  \alpha(G)$, it follows immediately that $G$ is not
  $\alpha$-vertex-stable. Otherwise, $G$ is
  $\alpha$-vertex-stable. Hence, this results in
  $\alpha$-$\textsc{VertexStability}$ belonging to $\Pol$ for
  co-graphs.~\eproofof{Theorem~\ref{thm:cograph-alpha-vertexstability}}

  \sproofof{Corollary~\ref{cor:co-graph-omega-vertex-stability}}
  Let $G \in \calC$ be a co-graph. Then its complement $\overline{G}$
  is a co-graph, too. Hence, we can exploit the fact that $\omega(G) =
  \alpha(\overline{G})$ holds and reuse the same idea as in
  \cref{thm:cograph-alpha-vertexstability} to decide whether $G$ is
  $\omega$-vertex-stable.~\eproofof{Corollary~\ref{cor:co-graph-omega-vertex-stability}}

  \sproofof{Proposition~\ref{prop:cograph-critical-vertex-subgraph}}
  We prove both cases separately.
  \begin{enumerate}
  \item If $G = G_1 \cup G_2$, it holds that $\chi(G) = \max \{
    \chi(G_1), \chi(G_2) \}$. Furthermore, if $u$ is $\chi$-critical for
    $G$, then it holds that $\chi(G - u) = \chi(G) - 1$. As we assume $u
    \in V(G_1)$, the removal of $u$ from $G$ only affects $G_1$, i.e., $G
    - u = (G_1 - u) \cup G_2$. Therefore, it follows that $\chi(G) =
    \chi(G_1) > \chi(G_2)$ must hold, as otherwise the removal of $u$
    would not affect $\chi(G)$. Consequently, $\chi(G_1 - u) = \chi(G_1) -
    1$ is true and $u$ is $\chi$-critical for $G_1$.
  \item If $G = G_1 + G_2$ holds, we have
    \begin{alignat*}{2}
      && \chi(G - u) & = \chi(G) - 1 \\
      \Rightarrow \quad && \chi((G_1 - u) + G_2) & = \chi(G_1) +
      \chi(G_2) - 1 \\
      \Rightarrow \quad && \chi(G_1 - u) + \chi(G_2) & = \chi(G_1)
      + \chi(G_2) - 1 \\
      \Rightarrow \quad && \chi(G_1 - u) & = \chi(G_1) - 1,
    \end{alignat*}
    such that $u$ is $\chi$-critical for $G_1$.~\eproofof{Proposition~\ref{prop:cograph-critical-vertex-subgraph}}
  \end{enumerate}

  \sproofof{Proposition~\ref{prop:co-graph-critical-vertices-critical-edge}}
  Let $G \in \calC$ be a co-graph and $e = \{ u, v \} \in E(G)$ an
  edge with two $\chi$-critical vertices $u,v \in V(G)$. First, we study
  the case that $G = G_1 + G_2$ as well as $u \in V(G_1)$ and $v \in
  V(G_2)$ holds. Afterwards, we explain how to generalize the proof.

  From the previous Proposition \ref{prop:cograph-critical-vertex-subgraph}
  we know that $u$ must be $\chi$-critical for $G_1$ and $v$
  $\chi$-critical for $G_2$. According to Observation 4 from
  \cite{fre-hem-rot:c:complexity-of-stability} there exists an optimal
  coloring $c_1 \colon V(G_1) \rightarrow \N$ for $G_1$, such that for
  all $\tilde{u} \in V(G_1) \setminus \{ u \}$ it holds that
  $c_1(\tilde{u}) \neq c_1(u)$. In other words, there is a coloring
  $c_1$ for $G_1$, such that $u$ is the only vertex in $G_1$ of its
  color. A similar, optimal coloring $c_2$ must exist for $G_2$ with
  respect to $v$. For the combined graph with $e$ removed, i.e., $G -
  e$, according to Observation 1 from
  \cite{fre-hem-rot:c:complexity-of-stability}, it must hold that
  $\chi(G - e) \in \{ \chi(G) - 1, \chi(G) \}$. Consequently, we can
  reuse $c_1$ and $c_2$ from $G_1$ and $G_2$, assuming distinct colors
  sets for $c_1$ and $c_2$, to obtain a legal coloring of $G$ with
  $\chi(G)$ colors. However, we can color $u$ in the same color
  $c_2(v)$, as $v$ is colored, and thus obtain a legal coloring for $G -
  e$ with $\chi(G) - 1$ colors. This is possible because
  \begin{enumerate}
  \item $u$ is the only vertex in $G_1$ colored in $c_1(u)$ by
    definition of $c_1$,
  \item no vertex $\tilde{u} \in V(G_1) \setminus \{ u \}$ is colored
    with $c_2(v)$, as $c_1(V(G_1)) \cap c_2(V(G_2)) = \emptyset$ holds, and
  \item $v$ is the only vertex in $G_2$ with this color, by definition
    of $c_2$, and there is no edge between $u$ and $v$.
  \end{enumerate}
  Consequently, after removing $e$ from $G$, we can color $G - e$ with
  one color less than before, such that $\chi(G - e) = \chi(G) - 1$
  holds and $e$ is $\chi$-critical.

  Initially, we assumed that $G = G_1 + G_2$ with $u \in V(G_1)$ and
  $v \in V(G_2)$ holds. If $G = G_1 \cup G_2$, there cannot exist any
  edge between vertices from $G_1$ and $G_2$. Hence, the only cases left
  are $G = G_1 + G_2$ or $G = G_1 \cup G_2$ with both vertices in $G_1$
  or $G_2$. Without loss of generality, let us assume that both vertices
  are in $G_1$. Following Proposition
  \ref{prop:cograph-critical-vertex-subgraph}, we know that both vertices
  are $\chi$-critical for $G_1$, as they are $\chi$-critical for $G$.
  When we can show that $e$ is $\chi$-critical for $G_1$, it immediately
  follows that $e$ is also $\chi$-critical for $G$. That is because if
  $G = G_1 + G_2$ and $e$ is $\chi$-critical for $G_1$, we have
  $\chi(G_1 - e) = \chi(G_1) - 1$, such that
  \begin{align*}
    \chi(G - e) = \chi(G_1 - e) + \chi(G_2) = \chi(G_1) - 1 +
    \chi(G_2) = \chi(G) - 1
  \end{align*}
  holds. If $G = G_1 \cup G_2$, there is one more argument to add. We
  know that $u$ and $v$ are $\chi$-critical for $G$ and
  $G_1$. Consequently, $\chi(G_1) > \chi(G_2)$ must hold, as
  otherwise, $u$ or $v$ cannot be $\chi$-critical for $G$, since
  $\chi(G) = \max \{ \chi(G_1), \chi(G_2) \}$ is true. But then, it is
  enough to show that $e$ is $\chi$-critical for $G_1$, since reducing
  $\chi(G_1)$ by one via removing $e$ also causes a reduction of
  $\chi(G)$ by one and hence, $e$ is $\chi$-critical for $G$, too.

  At some point, we must arrive in the case that one vertex is in
  $G_1$ and the other vertex is in $G_2$ and $G = G_1 + G_2$ holds,
  since the $+$-operation is the only possibility to add edges between
  vertices in co-graphs.~\eproofof{Proposition~\ref{prop:co-graph-critical-vertices-critical-edge}}

  \sproofof{Theorem~\ref{thm:co-graph-chi-stability}}
  Let $G \in \calC$ be a co-graph. We can compute $\chi(G)$
  efficiently and, according to Observation 1 in
  \cite{fre-hem-rot:c:complexity-of-stability}, for every edge $e \in
  E(G)$ and every vertex $v \in V(G)$ it holds that
  $\chi(G - e), \chi(G - v) \in \{ \chi(G) - 1, \chi(G) \}.$
  Thus, for every edge $e \in E(G)$, we proceed as follows to
  efficiently determine whether $e$ is $\chi$-critical or -stable for
  $G$: Denote $e = \{ u, v \}$ for $u,v \in V(G)$. Then it follows that
  $G - u$ and $G - v$ are induced subgraphs of $G - e$ and $G - e$ is a
  subgraph of $G$. According to the earlier referenced Observation 1,
  it must hold that
  \begin{align*}
    \underbrace{\chi(G - v), \chi(G - u)}_{\in \{ \chi(G)
    - 1, \chi(G) \}} \leq \chi(G - e) \leq \chi(G).
  \end{align*}
  Hence, if $\chi(G - v) = \chi(G)$ or $\chi(G - u) = \chi(G)$, which
  we can compute efficiently, it immediately follows that $\chi(G - e) =
  \chi(G)$. In other words, if $u$ or $v$ is $\chi$-stable, we know that
  $e$ must be $\chi$-stable, too.\footnote{This is in line with
    Observation 3 from \cite{fre-hem-rot:c:complexity-of-stability}.} If
  $u$ and $v$ are $\chi$-critical, it follows by Proposition
  \ref{prop:co-graph-critical-vertices-critical-edge} that $e$ is
  $\chi$-critical. Since we can determine for every node in $V(G)$
  efficiently, whether it is $\chi$-stable, we can also efficiently
  determine for every edge in $E(G)$ whether it is
  $\chi$-stable. Consequently, we can decide in polynomial time whether
  $G$ is $\chi$-stable. Thus $\chi$-$\textsc{Stability} \in \Pol$ for
  co-graphs follows.~\eproofof{Theorem~\ref{thm:co-graph-chi-stability}}

  \sproofof{Lemma~\ref{lem:co-graph:set-of-cliques}}
  For a co-graph $G \in \calC$, let us denote by $W(G)$ the set of all
  cliques of $G$ of size $\omega(G)$. Then the result easily follows by
  the recursive nature of co-graphs. To begin, if $G = (\{ v \},
  \emptyset)$, obviously it holds that $W(G) = \{\{ v \}\}$. If $G = G_1
  \cup G_2$, we have
  \begin{align*}
    W(G) =
    \begin{cases}
      W(G_1), & \omega(G_1) > \omega(G_2), \\
      W(G_2), & \omega(G_2) > \omega(G_1), \\
      W(G_1) \cup W(G_2), & \text{else}.
    \end{cases}
  \end{align*}
  If $G = G_1 + G_2$, we have
  \begin{align*}
    W(G) = \{ w_1 \cup w_2 \mid w_1 \in W(G_1), w_2 \in W(G_2) \}.
  \end{align*}
  Since we can compute a co-graph's co-tree as well as the size of its
  biggest clique(s) efficiently, it follows that the previously
  described algorithm to compute $W(G)$ can be executed in polynomial
  time, too.~\eproofof{Lemma~\ref{lem:co-graph:set-of-cliques}}

  \sproofof{Lemma~\ref{lem:co-graph-omega-reduce-by-one}}
  First of all, it is obvious that by removing a vertex or an edge
  from $G$ we cannot increase $\omega(G)$. Hence, $\omega(G - v),
  \omega(G - e) \leq \omega(G)$ holds. Furthermore, when we remove $v$
  from $G$, either $v$ is part of all cliques in $G$ of size
  $\omega(G)$, so that $\omega(G - v) = \omega(G) - 1$ holds or $v$ is
  not part of all of them, so that $\omega(G - v) = \omega(G)$
  holds. Generally speaking, by removing a vertex from $G$, we can
  either reduce a clique's size by one or leave it as it is. Now, let $e
  = \{u,v\} \in E(G)$ be an edge of $G$. Either $e$ is between two
  vertices of a clique in $G$ of size $\omega(G)$ or not. If that is the
  case, we reduce the clique's seize by one or leave it
  unchanged. Hence, by removing an edge from $G$, we either reduce
  $\omega(G)$ by one or do not alter it at all. Therefore, for all $v
  \in V(G)$ and $e \in E(G)$ it holds that $\omega(G - v), \omega(G - e)
  \in \{ \omega(G) - 1, \omega(G) \}$.~\eproofof{Lemma~\ref{lem:co-graph-omega-reduce-by-one}}

  \sproofof{Theorem~\ref{thm:co-graph-omega-stability}}
  Let $G \in \calC$ be a co-graph. By
  Theorem~\ref{thm:cograph-chromatic-number} we can compute $\omega$
  efficiently for $G$ and all induced subgraphs. In order to decide
  whether $G$ is $\omega$-stable, we proceed as follows for every edge
  $e = \{ u, v \}\in E(G)$:

  \textbf{Case 1:} $G = G_1 \cup G_2$ for two co-graphs $G_1,
  G_2$, and either $u,v \in V(G_1)$ or $u,v \in V(G_2)$ holds, since
  there are no edges between $G_1$ and $G_2$. Assume without loss of
  generality that $u,v \in V(G_1)$. As $\omega(G) = \max \{ \omega(G_1),
  \omega(G_2) \}$, we efficiently check whether $\omega(G_2) \geq
  \omega(G_1)$ holds. In this case, we know that $e$ cannot be critical
  to $G$, because even if $e$ would be $\omega$-critical to $G_1$, using
  Lemma~\ref{lem:co-graph-omega-reduce-by-one}, we still have
  $\omega(G - e) = \max \{ \omega(G_1 - e), \omega(G_2) \} = \max
    \{ \omega(G_1) - 1, \omega(G_2) \} = \omega(G_2)$.
  Otherwise, if $\omega(G_1) > \omega(G_2)$ holds, we study whether $e$ is
  $\omega$-critical for $G_1$ by recursively selecting the appropriate
  case, this time with $G_1$ as $G$. This is sufficient because if $e$ is
  $\omega$-critical for $G_1$, it is also $\omega$-critical for $G$.

  \textbf{Case 2:} $G = G_1 + G_2$ and $u,v \in V(G_1)$ or $u,v
  \in V(G_2)$. In this case, it is sufficient to check whether $e$ is
  $\omega$-critical for the partial graph, i.e., $G_1$ or $G_2$,
  containing $u$ and $v$. That is because $\omega(G) = \omega(G_1) +
  \omega(G_2)$ holds and so, if $e$ is $\omega$-critical for one of the
  two partial graphs, $e$ is also critical for $G$. Once again, we
  check this by recursively applying the appropriate case for the
  corresponding partial graph.

  \textbf{Case 3:} $G = G_1 + G_2$ and $u,v$ are in different
  partial graphs. Assume that $u \in V(G_1)$ and
  $v \in V(G_2)$ holds. Now, in order for $e$ to be $\omega$-critical,
  there must exist only one clique $V' \subseteq V(G_1)$ with
  $\omega(G_1) = |V'|$ as well as $u \in V'$ and only one clique $V''
  \subseteq V(G_2)$ with $\omega(G_2) = |V''|$ and $v \in V''$. We
  can check both conditions efficiently using
  Lemma~\ref{lem:co-graph:set-of-cliques}. If this
  is the case, then all other cliques in $G_1$ are strictly smaller than
  $V'$ and all other cliques in $G_2$ are strictly smaller than
  $V''$. Hence, the only clique of size $\omega(G)$ in $G$ is $V' \cup
  V''$, containing $u$ and $v$. Removing $e = \{u,v\}$ from $G$ causes
  $\omega(G)$ to be reduced by one since there is only one clique of
  size $\omega(G)$ in $G$, and afterwards, it is missing the edge $e$ in $G -
  e$. Therefore, only in this case $e$ is $\omega$-critical.

  The number of recursive calls is limited by $\lceil \log(|V(G)|)
  \rceil$, since every inner node of $G$'s co-expression combines at
  least two nodes. Every case can be computed efficiently, such that
  we can determine for a co-graph $G$ in time in $\calO(|E(G)| \cdot
  \log(|V(G)|) \cdot |V(G)|^c)$ for some $c \in \N$ whether $G$ is
  $\omega$-stable. Consequently, $\omega$-$\textsc{Stability}$ is in
  $\Pol$ for all co-graphs.~\eproofof{Theorem~\ref{thm:co-graph-omega-stability}}

  \sproofof{Corollary~\ref{cor:co-graph-alpha-stability}}
  Let $G \in \calC$ be a co-graph. Then $\overline{G}$ is a co-graph,
  such that $\alpha(G) = \omega(\overline{G})$ holds. If $\overline{G}$
  is $\omega$-stable, which we can determine efficiently by
  \cref{thm:co-graph-omega-stability}, it follows immediately that $G$
  is $\alpha$-stable. Hence, we can efficiently determine whether a
  co-graph $G$ is $\alpha$-stable, such that
  $\alpha$-$\textsc{Stability}$ is in $\Pol$ for
  co-graphs.~\eproofof{Corollary~\ref{cor:co-graph-alpha-stability}}

  \sproofof{Corollary~\ref{cor:co-graph-alpha-beta-unfrozenness}}
  Let $G \in \calC$ be a co-graph. Co-graphs are closed under
  complements, so $\overline{G}$ is a co-graph as well and we can
  compute $\overline{G}$ from $G$ efficiently. According to \cref
  {thm:co-graph-omega-stability}, we can check in time polynomial in
  $|G|$ whether $\overline{G}$ is $\omega$-stable. Using Proposition 5
  2. from \cite{fre-hem-rot:c:complexity-of-stability}, we immediately
  know that $G$ is $\beta$- and $\alpha$-unfrozen if $\overline{G}$ is
  $\omega$-stable. Hence, both problems, $\beta$- and
  $\alpha$-$\unfrozenness$ are in $\Pol$ for all
  co-graphs.~\eproofof{Corollary~\ref{cor:co-graph-alpha-beta-unfrozenness}}

  \sproofof{Corollary~\ref{cor:co-graph-omega-unfrozenness}}
  From \cref{cor:co-graph-alpha-stability} we know that we can
  efficiently decide for a co-graph whether it is
  $\alpha$-stable. Hence, let $G \in \calG$ be a co-graph. Consequently,
  $\overline{G}$ is a co-graph, too, and we can calculate for
  $\overline{G}$ in time polynomial in $|G|$ whether it is
  $\alpha$-stable. Applying Proposition 5 1. from
  \cite{fre-hem-rot:c:complexity-of-stability}, we know that if
  $\overline{G}$ is $\alpha$-stable, it follows that $G$ is
  $\omega$-unfrozen. Therefore, we can efficiently calculate whether $G$
  is
  $\omega$-unfrozen.~\eproofof{Corollary~\ref{cor:co-graph-omega-unfrozenness}}

  \sproofof{Theorem~\ref{thm:co-graph-chi-unfrozenness}}
  Let $G \in \calG$ be a co-graph and $e = \{u,v\} \in
  \overline{E}(G)$ a nonedge of $G$. Since $G$ has at least two
  vertices, $u$ and $v$, either $G = G_1 + G_2$ or $G = G_1 \cup G_2$ for two
  co-graphs $G_1, G_2$ holds. We handle both cases separately:
  \begin{enumerate}
  \item If $G = G_1 + G_2$ is true, then $e$ must belong either to
    $\overline{E}(G_1)$ or to $\overline{E}(G_2)$, since $V(G_1)
    \times V(G_2) \subseteq E(G)$, such that $\overline{E}(G) =
    \overline{E}(G_1) \cup \overline{E}(G_2)$. Without loss of
    generality assume that $e \in \overline{E}(G_1)$ holds. If $e$ is
    $\chi$-unfrozen for $G_1$, i.e., $\chi(G_1 + e) = \chi(G_1)$, then
    $e$ is $\chi$-unfrozen for $G$, since $\chi(G + e) = \chi(G_1 + e) +
    \chi(G_2) = \chi(G_1) + \chi(G_2) = \chi(G)$ follows. Contrarily, if $e$ is
    $\chi$-frozen for $G_1$, i.e., $\chi(G_1 + e) = \chi(G_1) + 1$,
    then $e$ is $\chi$-frozen for $G$ as well, as $\chi(G + e) =
    \chi(G_1 + e) + \chi(G_2) = \chi(G_1) + 1 + \chi(G_2) = \chi(G) +
    1$ holds. Hence, it is enough to determine whether $e$ is
    $\chi$-unfrozen or -frozen for $G_1$ and we can follow the
    argumentation of this proof recursively for $G_1$.
  \item If $G = G_1 \cup G_2$ is true, $e$ can belong to
    $\overline{E}(G_1), \overline{E}(G_2)$ or $\overline{E}(G)$. We
    split this into two sub-cases:
    \begin{enumerate}
    \item If $e \in \overline{E}(G_1)$ or $e \in \overline{E}(G_2)$,
      we proceed as follows. Without loss of generality assume $e \in
      \overline{E}(G_1)$. Since $\chi(G) = \max \{ \chi(G_1),
      \chi(G_2) \}$ holds, an increase of $\chi(G_1)$ affects $\chi(G)$
      only if $\chi(G_1) \geq \chi(G_2)$. Otherwise, $e$ is
      $\chi$-unfrozen for $G$ (but not necessarily for $G_1$). If
      $\chi(G_1) \geq \chi(G_2)$, then it holds that if $e$ is
      $\chi$-unfrozen for $G_1$, it follows that $e$ is
      $\chi$-unfrozen for $G$, since $\chi(G + e) = \max \{ \chi(G_1 +
      e) , \chi(G_2) \} = \chi(G_1 + e) = \chi(G_1) = \chi(G)$ is
      true. Similarly, if $e$ is $\chi$-frozen for $G_1$, it follows
      that $e$ is $\chi$-frozen for $G$, since $\chi(G + e) = \max \{
      \chi(G_1 + e), \chi(G_2) \} = \chi(G_1 + e) = \chi(G_1) + 1 =
      \chi(G) + 1$. Consequently, it is enough to determine whether
      $e$ is $\chi$-unfrozen or -frozen for $G_1$ and we can follow
      the argumentation of this proof recursively for $G_1$.
    \item If $e \in \overline{E}(G) \setminus (\overline{E}(G_1) \cup
      \overline{E}(G_2))$, then $u \in V(G_1)$ and $v \in V(G_2)$
      follows. Now, if $\chi(G_1) = \chi(G_2) = 1$ is true, it follows
      that $e$ is $\chi$-frozen for $G$, since $\chi(G + e) = 1 + 1 =
      2 > 1 = \max\{ \chi(G_1), \chi(G_2)\}  =\chi(G)$. Contrarily, if
      $\chi(G_1) > 1$ or $\chi(G_2) > 1$, it follows that $e$ is
      $\chi$-unfrozen for $G$ since $G_1$ and $G_2$ share no edge but $e$.
      Because of that we can arrange the colors for $V(G_1)$ and $V(G_2)$ in
      such a way that both vertices incident to $e$ have different
      colors, resulting in $\chi(G + e) = \chi(G)$.
    \end{enumerate}
  \end{enumerate}
  Following these cases, we can efficiently determine for every
  nonedge $e \in \overline{E}(G)$ whether it is $\chi$-frozen or
  -unfrozen for $G$, resulting in $\chi$-$\unfrozenness \in \Pol$ for
  all co-graphs.~\eproofof{Theorem~\ref{thm:co-graph-chi-unfrozenness}}

\section{Gallai's Theorem}

For the sake of self-containment, we here state Gallai's
theorem~\cite{gal:j:extreme-punkt-kantenmengen}, which is used to
obtain several of our results, and we also provide its proof.

\begin{theorem}[Gallai's theorem]
  \label{thm:gallai}
  For every graph $G \in \calG$, it holds that
  \begin{align*}
    |V(G)| = \alpha(G) + \beta(G).
  \end{align*}
\end{theorem}

\begin{proofs}
  Let $V' \subseteq V(G)$ be a vertex cover for $G$ of size $\beta(G)$
  and assume that there are two vertices $u,v \in V \setminus V'$ which
  are adjacent, i.e., $\{u,v\} \in E(G)$. This contradicts the fact that
  $V'$ is a vertex cover for $G$, as $\{u,v\}$ would not be covered by
  $V'$. Consequently, $V \setminus V'$ must be an independent set for
  $G$ and we obtain
  \begin{align}
    \alpha(G) \geq |V(G)| - \beta(G). \label{eq:gallais-theorem-1}
  \end{align}
  Let $V'' \subseteq V(G)$ be an independent set for $G$ of size
  $\alpha(G)$. For every edge $\{u,v\} \in E$ it must hold that either
  $u$ or $v$ is not in $V''$, as this would contradict the fact that
  $V''$ is an independent set. Hence, $V \setminus V''$ must be a vertex
  cover for $G$ and we obtain
  \begin{align}
    \beta(G) \leq |V(G)| - \alpha(G). \label{eq:gallais-theorem-2}
  \end{align}
  \cref{eq:gallais-theorem-1} yields $\alpha(G) + \beta(G) \geq |V(G)$
  and \cref{eq:gallais-theorem-2} yields $\alpha(G) + \beta(G) \leq
  |V(G)$, such that we obtain
  \begin{align*}
    |V(G)| = \alpha(G) + \beta(G).
  \end{align*}
This completes the proof.~\end{proofs}

\end{document}